\documentclass[%
 reprint,
 superscriptaddress,
 amsmath,amssymb,
 aps,
 prl,
 longbibliography
]{revtex4-2}

\usepackage{enumitem}
\usepackage{etoolbox}
\usepackage[
 colorlinks=true,
 citecolor=blue,
 linkcolor=blue
]{hyperref}
\usepackage[pdftex]{graphicx} 

\makeatletter
\newcommand{\DisableTOC}{%
  \let\saved@addcontentsline\addcontentsline
  \renewcommand{\addcontentsline}[3]{}%
}
\newcommand{\EnableTOC}{%
  \let\addcontentsline\saved@addcontentsline
}
\makeatother

\usepackage{braket}
\usepackage{times,multirow,amsfonts,bm,xspace,pifont,mathtools}
\usepackage{soul}
\usepackage{ulem}
\usepackage{mathrsfs}
\usepackage{array}
\usepackage[T1]{fontenc}
\usepackage[utf8]{inputenc}

\newcommand{\LastBibLabel}{LastMainBibItem}
\makeatletter
\patchcmd{\NAT@bibsetnum}{LastBibItem}{\LastBibLabel}{}{
  \PackageError{arxiv}{Could not separate bibliography widths}{}
}
\patchcmd{\endthebibliography}{LastBibItem}{\LastBibLabel}{}{
  \PackageError{arxiv}{Could not separate bibliography labels}{}
}
\makeatother

\begin{document}

\newcommand{\rr}{{\bm r}}
\newcommand{\q}{{\bm q}}
\renewcommand{\k}{{\bm k}}
\newcommand*\YY[1]{\textcolor{blue}{#1}}
\newcommand{\YYS}[1]{\textcolor{blue}{\sout{#1}}}
\newcommand*\YYWC[1]{\textcolor{green}{#1}}
\newcommand{\TK}[1]{{#1}}
\newcommand*\TKS[1]{\textcolor{red}{\sout{#1}}}

\newcommand{\AD}[1]{{\color{magenta}{#1}}}
\newcommand*\ADS[1]{\textcolor{magenta}{\sout{#1}}}

\newcommand*\BdG{{\rm BdG}}

\title{Quantum geometric ferromagnetism by singular saddle point
}

\author{Taisei Kitamura}
\email[]{taisei-kitamura@g.ecc.u-tokyo.ac.jp}
\affiliation{Department of Physics, Graduate School of Science, Kyoto University, Kyoto 606-8502, Japan}
\affiliation{RIKEN Center for Emergent Matter Science (CEMS), Wako 351-0198, Japan}
\affiliation{Department of Physics, Graduate School of Science, The University of Tokyo, Tokyo 113-0033, Japan}

\author{Hiroki Nakai}
\affiliation{Graduate School of Arts and Sciences, University of Tokyo, Tokyo 153-8902, Japan}

\author{Akito Daido}
\affiliation{Department of Physics, Graduate School of Science, Kyoto University, Kyoto 606-8502, Japan}


\author{Youichi Yanase}
\affiliation{Department of Physics, Graduate School of Science, Kyoto University, Kyoto 606-8502, Japan}

\date{\today}

\begin{abstract}
We propose ferromagnetism that occurs in electrons at a saddle point with band touching, which we call the \textit{singular saddle point}. 
At the singular saddle point, the divergent quantum metric induces ferromagnetic correlation, and the logarithmic divergence of the density of states ensures ferromagnetism within Stoner theory. This is a prototypical example of \textit{quantum geometric ferromagnetism}.
The two-dimensional $t_{2g}$-orbital model accommodates the ferromagnetism by this mechanism, which is continuously connected to the exactly proven flat-band ferromagnetism.
\end{abstract}

\maketitle

\DisableTOC

\textit{Introduction---}Itinerant ferromagnetism has long been a central topic in condensed matter physics~\cite{Moriya1985}. A canonical example is the strongly correlated $3d$-electron systems, such as Fe, Co, and Ni, which exhibit ferromagnetism as their three-dimensional bands are partially filled with electrons~\cite{Terakura1982}. Heavy fermion materials provide another platform for studying itinerant ferromagnetism, and the relationship with unconventional superconductivity has been intensively studied~\cite{Saxena2000,Aoki2001,Huy2007,Aoki2019}. 
In contrast, two-dimensional (2D) magnetism in van der Waals materials has recently attracted much attention with potential applications for spintronics~\cite{Burch2018,Gong2019,Wang2022}, and itinerant 2D ferromagnets have been discovered~\cite{Gong2017,Huang2017,Deng2018,Fei2018,May2016,May2019,Nakano2019,Seo2020}.

To reveal the origin of itinerant ferromagnetism, intensive theoretical studies have been carried out~\cite{Hubbard1963,Gutzwiller1963,Kanamori1963,Nagaoka,Moriya1973,Tasaki1998}. 
Based on Stoner's mean-field theory~\cite{Stoner1936}, the large density of states (DOS) is favorable for ferromagnetism; this concept is extended by a framework beyond mean-field theory by Kanamori~\cite{Kanamori1963}. 
An extreme case is flat-band systems, in which the DOS is infinite, and the ferromagnetic ground state is rigorously proven~\cite{Mielke1993,Tasaki1998,Mielke1999,Katsura2010,Tasaki2020} for example in the Lieb lattice~\cite{Lieb1989}, the kagome lattice by Mielke~\cite{Mielke1991,Mielke1992}, and the Tasaki lattice~\cite{Tasaki1992}. In some models, ferromagnetism remains stable against perturbations that break the flatness of the band
~\cite{Tasaki1995,Tanaka2001,Tanaka2003,Ueda2004,Lu2009,Tanaka2018,Tamura2019}. 
Although these theories provide insight into the origin of itinerant ferromagnetism, realizing materials with large DOS that resemble flat-band systems is challenging. 
Furthermore, the relation between exact theories of ferromagnetism and real ferromagnetic materials is elusive. 

These issues are particularly crucial in the study of 2D systems. 
The recent discovery of itinerant ferromagnetism in 2D van der Waals materials~\cite{Fei2018,Deng2018} paves the way for exploring functional quantum materials and highlights the need for guiding principles in the search for a broad class of 2D ferromagnets.
In 2D systems, a saddle point in the band dispersion gives rise to a logarithmic divergence of the DOS, namely a van Hove singularity, which can be advantageous for ferromagnetism~\cite{Honerkamp2001,Yu-Irkhin2001,Katanin2003}; stronger, e.g. power-law, divergences can occur at higher-order saddle points, often referred to as higher-order van Hove singularities~\cite{Yuan2019-hm,Classen2025-kl}.
However, although the saddle point is an ubiquitous feature in the square lattice, the ground state is mostly antiferromagnetic~\cite{Yanase2003}.
Thus, despite the general insights obtained from theories, the prediction of 2D ferromagnetic materials remains a challenging task.

\begin{figure}[tbp]
  \includegraphics[width=1.0\linewidth]{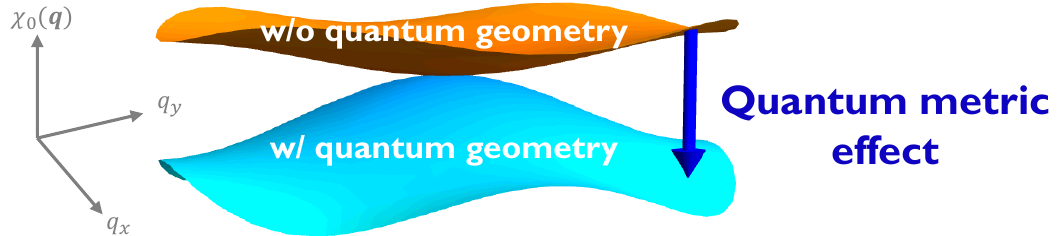}
  \centering
  \caption{Illustration of the mechanism of QGFM. Total spin susceptibility (blue surface) shows a peak at $\bm q =0$ indicating ferromagnetic correlation due to the quantum metric,   although spin susceptibility without quantum geometry (orange surface) shows antiferromagnetic correlation.}
  \label{fig:schematic}
\end{figure}

Recently, another route to ferromagnetism has been proposed by theory using quantum geometry~\cite{Kitamura2024}, inspired by the exploration of various phenomena due to quantum geometry in condensed matter physics~\cite{Resta2011,Cheng2013,Rossi2021,Torma2022,Torma2023,Yu2024}.
In general, the spin susceptibility of noninteracting systems, $\chi_0(\bm q)$ with momentum $\bm q$, can be divided into the quantum geometric term and the energy-dispersion term.
The quantum geometric term near $\bm q= 0$ arises from the quantum metric that represents the distance between two adjacent Bloch states~\cite{Kitamura2024}.
In Ref.~\onlinecite{Kitamura2024}, it has been shown that the quantum metric generally favors ferromagnetic correlation by suppressing antiferromagnetic correlation, as schematically illustrated in Fig.~\ref{fig:schematic}. 
Therefore, in systems with significant quantum geometry, ferromagnetism can be triggered by the Coulomb interaction, and we call it the \textit{quantum geometric ferromagnetism} (QGFM).

Recent studies have also revealed the relationship between quantum geometry and various classes of magnetism~\cite{Herzog2022,Yu2024,Heinsdorf2024,Wu2020,Kang2024,Zhang2025} and unconventional superconductivity~\cite{Kitamura2022,Jiang2023,Chen2023,Kitamura2023,Sun2024,Dunbrack2025,Shavit2024,Jahin2025}.
After the proposal of QGFM~\cite{Kitamura2024}, the quantum geometric origin of magnetism in a perfect flat band has also been studied and clarified~\cite{Kang2024,Zhang2025}.
Thus, the interplay of quantum geometry and electron correlation is now expected to be a new paradigm of quantum many-body physics.

In this Letter, we establish a theoretical framework for a prototypical class of QGFM and provide a guideline for searching the platform of 2D ferromagnetism.
A key mechanism relies on the \textit{singular saddle point}, where the band touching occurs at the saddle point.
When the singular saddle point exists at the Fermi energy, both the DOS and quantum geometry diverge, leading to the emergence of QGFM.
The singular saddle point ubiquitously emerges at the high-symmetry point with $C_4$ symmetry in the 2D multi-band models. 
We show that the 2D $t_{2g}$-orbital model is an example of QGFM originating from the singular saddle point.
Analyzing this model, we show a close relation between exactly proven flat-band ferromagnetism and QGFM, providing a link between exact theories and real materials.

 \begin{figure}[tbp]
  \includegraphics[width=1.0\linewidth]{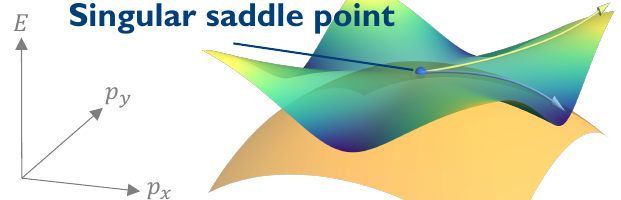}
  \centering
  \caption{Schematic illustration of the singular saddle point. The blue sphere highlights the band touching point. We set the origin of the momentum at the band touching point, that is, $\bm p = 0$. 
  The blue and yellow arrows represent the downward and upward dispersion of the upper band along the $p_{x}$ and $p_x = p_y$ directions, respectively.}
  \label{fig:schematic_spp}
\end{figure}

\textit{Singular saddle point---}The singular saddle point proposed in this Letter is schematically illustrated in Fig.~\ref{fig:schematic_spp}.
Let us consider the high-symmetry point where band touching is protected by a certain symmetry. 
In this case, one of the degenerate bands can exhibit an energy dispersion that is opposite between the $p_x$ (and $p_y$) direction and the diagonal $p_x = p_y$ direction. 
This contrasts to conventional saddle points, where the sign of effective mass is different between two orthogonal directions, such as along the $p_x$ and $p_y$ axes.

Here, we show that the $C_4$-symmetric multi-band systems can host singular saddle points. To be specific, in this Letter, we consider 2D systems with two-fold band degeneracy protected by the $C_4$ rotation symmetry~\cite{symmetry_comment}.
The characteristic feature of the singular saddle point is modeled by the $kp$-perturbation Hamiltonian for the two-band model, 
$
    H_{\rm kp}(\bm p) = h_0(\bm p)\sigma_0 + \bm h(\bm p)\cdot \bm \sigma
$,
with the unit matrix $\sigma_0$ and the Pauli matrices $\bm \sigma = (\sigma_x,\sigma_y,\sigma_z)$.
The energy dispersion is obtained as $\epsilon_\pm(\bm p) = h_0(\bm p) \pm\vert\bm h(\bm p)\vert$.
Since the two bands are degenerate at $\bm p = 0$, $\bm h(0) = 0$ is satisfied, and we can show that the quantum metric at $\bm p = 0$ diverges.
Without loss of generality,
we set the origin of the energy to be $\epsilon_\pm(\bm 0)$, i.e., $h_0(\bm 0) = 0$.
As a result, the Hamiltonian $H_{\rm kp}(\bm p)$ contains only terms proportional to the lowest order in $\bm p$, which is assumed to be the second order unless otherwise stated.

The condition of the singular saddle point is obtained by analyzing the Hamiltonian on high-symmetry lines. 
The minimal Hamiltonian with the band touching protected by the $C_4$ rotation symmetry
is described by
$[h_0(\bm p),h_x(\bm p),h_y(\bm p),h_z(\bm p)] = [(p_x^2+p_y^2)/2M, p_xp_y/M_{xy}, 0, (p_x^2-p_y^2)/2M_{xx}]$ or its unitary equivalents.
Here, $M_{xy},M_{xx} \ge 0,$ and $M$ depend on the model details.
Comparing the sign of the effective masses at $\bm p =0$ along the high-symmetry directions $p_{y(x)} = 0$ and $p_x = p_y$, 
we find that a singular saddle point appears when $1/M_{xy} > 1/|M| > 1/M_{xx}$ or $1/M_{xy} < 1/|M| < 1/M_{xx}$. 
The lower band $\epsilon_-(\TK{\bm p})$ can host a saddle point when $1/M > 0$, while the upper band $\epsilon_+(\TK{\bm p})$ can do it when $1/M < 0$.

Next, we show that the DOS shows a logarithmic divergence at the energy of the singular saddle point.
The DOS can be estimated for the approximated band dispersion,
$
    \epsilon_{\pm}^{\rm eff}(\bm p) = 
    (p_{x}^2+p_{y}^2)/2m^{\pm}
  \pm \vert p_xp_y\vert /m_{xy}
$ 
with $1/m^{\pm} = 1/M \pm 1/M_{xx}$ and $1/m_{xy} =  1/M_{xy} -  1/M_{xx}$, 
which reproduces $\epsilon_\pm(\bm p)$
on the high-symmetry lines and smoothly complements between them.
The DOS of the band $\epsilon_\pm^{\rm eff}(\bm p)$, that is, $D_\pm(\varepsilon)$ is summarized in Table~\ref{table:DOS}~\cite{Appendix}. 
For the band with a saddle point, the DOS is the sum of the logarithmically divergent term $D_{\rm log}(\varepsilon) = -A\ln\vert\varepsilon\vert+B$ and the asymmetric term proportional to the step function $\theta(\pm \varepsilon)$ with a coefficient $-D_{\rm asym}$. 
Thus, the singular saddle point leads to a simultaneous divergence in the quantum metric and the DOS.
In contrast, the band without a saddle point shows an asymmetric constant DOS, $D_\pm(\varepsilon) = \theta(\pm \varepsilon) D_{\rm pb}$, as in the 2D parabolic band.

Let us discuss the differences between the conventional and singular saddle points. 
Because the sign of the effective mass is different in two orthogonal directions, conventional saddle points are prohibited from appearing at $C_4$-symmetric points, and instead appear at multiple momenta that are related to each other by the $C_4$ symmetry.
Typical examples are $X$ and $Y$ points in the 2D square lattice model. In the presence of such conventional saddle points, the antiferromagnetic correlation develops in many cases due to nesting between multiple saddle points~\cite{Yanase2003}.
In contrast, the singular saddle point can appear alone at a high-symmetry point such as the $\Gamma$ and M points, because the singular band dispersion can respect the $C_4$
symmetry, for which band touching is needed.
The presence of a single saddle point is expected to be advantageous for ferromagnetism because of the singular DOS and avoided coupling of multiple saddle points.



\renewcommand{\arraystretch}{1.1}
\begin{table}[tbp]
 \caption{DOS of the bands $\epsilon_\pm^{\rm eff}(\TK{\bm p})$, that is $D_\pm(\varepsilon)$, for each condition realizing a singular saddle point.}
 \label{table:DOS}
 \centering
  \resizebox{\columnwidth}{!}{%
  \begin{tabular}{ccc}
   \hline
   Condition& $D_+(\varepsilon)$ & $D_-(\varepsilon)$\\
   \hline \hline
    $\frac{1}{M_{xy}}>\frac{1}{M} > \frac{1}{M_{xx}}$ &
    $\theta(\varepsilon)D_{\rm pb}$ &
    $D_{\rm log}(\varepsilon) - \theta(\varepsilon)D_{\rm asym}$\\
    $\frac{1}{M_{xy}}<\frac{1}{M} < \frac{1}{M_{xx}}$ &
    $\theta(\varepsilon)D_{\rm pb}$ &
    $D_{\rm log}(\varepsilon) - \theta(-\varepsilon)D_{\rm asym}$
    \\
    $\frac{1}{M_{xy}}>\frac{-1}{M} > \frac{1}{M_{xx}}$ & $D_{\rm log}(\varepsilon) - \theta(-\varepsilon)D_{\rm asym}$ & $\theta(-\varepsilon)D_{\rm pb}$\\
    $\frac{1}{M_{xy}}<\frac{-1}{M} < \frac{1}{M_{xx}}$ & $D_{\rm log}(\varepsilon) - \theta(\varepsilon)D_{\rm asym}$&
    $\theta(-\varepsilon)D_{\rm pb}$\\
   \hline
  \end{tabular}%
  }
\end{table}

\textit{Quantum geometric ferromagnetism---}When the Fermi energy is located on the singular saddle point,
i.e., $\mu = 0$, the logarithmic divergence of the DOS is expected to favor ferromagnetism. However, singularity in the DOS is not a sufficient condition for ferromagnetism, and the effect of quantum geometry plays an essential role. 
To see this, we introduce a criterion for ferromagnetic correlation defined by the curvature of spin susceptibility, $\chi_{\rm c}^{i j} = \lim_{\bm q\rightarrow 0}\partial_{q_i}\partial_{q_j}\chi_0(\bm q)$. We can impose $\chi_{\rm c}^{xx} = \chi_{\rm c}^{yy}$ and $\chi_{\rm c}^{xy}=0$ in $C_4$-symmetric systems.
Although $\chi_{\rm c}^{xx}>0$ rules out the ferromagnetic correlation, $\chi_{\rm c}^{xx}<0$ is compatible with the ferromagnetic correlation, because it indicates that $\bm{q}=0$ is a local maximum of $\chi_0(\bm{q})$ (see Fig.~\ref{fig:schematic}).
In Ref.~\onlinecite{Kitamura2024},
the curvature $\chi_{\rm c}^{ij}$ is shown to be divided into two terms as $\chi_{\rm c}^{i j}=\chi_{\rm geom}^{ij} + \chi_{\rm mass}^{ij}$.
The quantum geometric term $\chi_{\rm geom}^{ij}$ includes the quantum metric, and $\chi_{\rm mass}^{ij}$ is the effective mass term determined by the band dispersion. The formulas for $\chi_{\rm geom}^{ij}$ and $\chi_{\rm mass}^{ij}$ are given in Appendix~\cite{Appendix}. 

To illustrate the effect of quantum geometry, let us neglect quantum geometry and consider an effective Hamiltonian, $H_{\rm eff}(\bm p) = {\rm diag}[\epsilon_+^{\rm eff}(\bm p),\epsilon_-^{\rm eff}(\bm p)]$. In this case, $\chi^{ij}_{\rm c}$ is given only by the effective mass term, which is divided into contributions from the two bands as $\chi_{\rm mass}^{ij} = \chi_{\rm mass:sp}^{ij} + \chi_{\rm mass:pb}^{ij}$. 
The band hosting the singular saddle point gives
$\chi_{\rm mass:sp}^{xx} = D_{\rm asym}/24\vert m^{\pm}\vert T >0$~\cite{Appendix}, which shows positive $1/T$ divergence~\cite{T_dive_comment}.
Therefore, ferromagnetic correlation is prohibited when we consider only the singular band with a saddle point. 
The other band with nearly parabolic dispersion gives $\chi_{\rm mass:pb}^{xx}  = -D_{\rm pb}/24 \vert m^\pm\vert T<0$ and negative $1/T$ divergence favors ferromagnetic correlation~\cite{Appendix}. Thus, the two bands compete in the effective mass term, and the ferromagnetic correlation can be suppressed by the effective mass term.
In fact, in the later analysis of the $t_{2g}$-orbital model, $\chi_{\rm mass:sp}^{ij}$ is shown to overcome $\chi_{\rm mass:pb}^{ij}$.
In this case, ferromagnetic correlation is forbidden if the effects of quantum geometry are absent.

Now we show that the quantum geometric contribution can lead to ferromagnetism. For analytic calculations, we adopt the polar coordinates $(p_{x}, p_{y}) = p (\cos\theta, \sin\theta)$, and analyze the generic two-band $kp$ Hamiltonian $H_{\rm kp}(\bm p)$ on the $n$-th order of $\bm p$.
Leaving out the terms of $\mathcal{O}(T^0)$, the following formula is obtained~\cite{Appendix},
\begin{align}    
	\chi_{\rm geom}^{ij} &=
	\dfrac{1}{4\pi^2}\int_{0}^{2\pi}d\theta
	\dfrac{g^{ij}(\theta)}{2nT}\left[\dfrac{1}{2}+I(\theta)\right],~\label{eq:geom_theta}
\end{align}
where $g^{ij}(\theta) = \TK{p^2\partial_{p_i}\hat{\bm h}(\theta)\cdot\partial_{p_j}\hat{\bm h}(\theta)}$ 
is the dimensionless quantum metric defined by the quantum metric multiplied by $p^2$.
Defining $h_0(\theta) = h_0(\bm p)/p^n$ and $\bm h(\theta) = \bm h(\bm p)/p^n$, we denote $\hat{\bm h}(\theta) = \bm h(\theta)/\vert\bm h(\theta)\vert$. 
The quantum geometric term $\chi_{\rm geom}^{ij}$ also diverges as $1/T$~\cite{T_dive_comment}.
At low temperatures, $I(\theta)$ is reduced to~\cite{Appendix},
\begin{eqnarray}
    I(\theta) = -\dfrac{h_0(\theta)}{4\vert \bm h(\theta)\vert}
	\ln\left\vert \dfrac{ h_0(\theta) + \vert \bm h(\theta)\vert}{h_0(\theta) - \vert \bm h(\theta)\vert}\right\vert.~\label{eq:zero_temp_I}
\end{eqnarray}
This formula can be applied to general cases with band degeneracy protected by symmetries including those other than the $C_4$ symmetry.
Because $I(\theta) < 0$, the second term in Eq.~\eqref{eq:geom_theta} favors ferromagnetic correlation.
In Eq.~\eqref{eq:zero_temp_I}, we see that the $\theta$-resolved geometric contribution shows a negative logarithmic divergence for $\theta$ on the Fermi surface, where $h_0(\theta) - \vert \bm h(\theta)\vert = 0$ or $h_0(\theta) + \vert \bm h(\theta)\vert = 0$. 

Let us again focus on the model with the band touching protected by the $C_4$ rotation symmetry and assume $n=2$.
When a singular saddle point appears, 
the sign of $h_0(\theta)\pm\vert\bm h(\theta)\vert$ changes eight times by changing $\theta$ from $0$ to $2\pi$, indicating eight Fermi surfaces. 
The presence of eight nodes in $h_0(\theta)\pm\vert\bm h(\theta)\vert$ makes the quantum geometric term $\chi_{\rm geom}^{xx}$ largely negative, which can induce ferromagnetic correlation by overcoming the potentially positive effective mass term $\chi_{\rm mass}^{xx}$. 
For the on-site Coulomb interaction $U$, the Stoner criterion for itinerant magnetism is given by $U\chi_0(\bm q)/2 > 1$~\cite{Nogaki2024}.
As we discussed above, the bare spin susceptibility can show a maximum at $\bm q =0$ due to the quantum geometric contribution, and the logarithmically divergent DOS leads to divergent ferromagnetic susceptibility $\chi_0(0)$ at low temperatures. 
Thus, the electron systems with a singular saddle point naturally satisfy the Stoner criterion for itinerant ferromagnetism.

\textit{Two-dimensional $t_{2g}$-orbital model---}To demonstrate QGFM by the singular saddle point, we analyze the 2D $t_{2g}$-orbital model.
Considering prototypical strongly correlated systems, such as $3d$ electron systems, we take into account the $t_{2g}$ orbitals. In tetragonal systems with $C_4$ symmetry, the $d_{xz}$ and $d_{yz}$ orbitals are entangled with each other, while the other $d_{xy}$ orbital can be separated from the other orbitals in energy. Therefore, we study the two-orbital models for the $d_{xz}$ and $d_{yz}$ orbitals.

\begin{figure}[tbp]
  \includegraphics[width=1.0\linewidth]{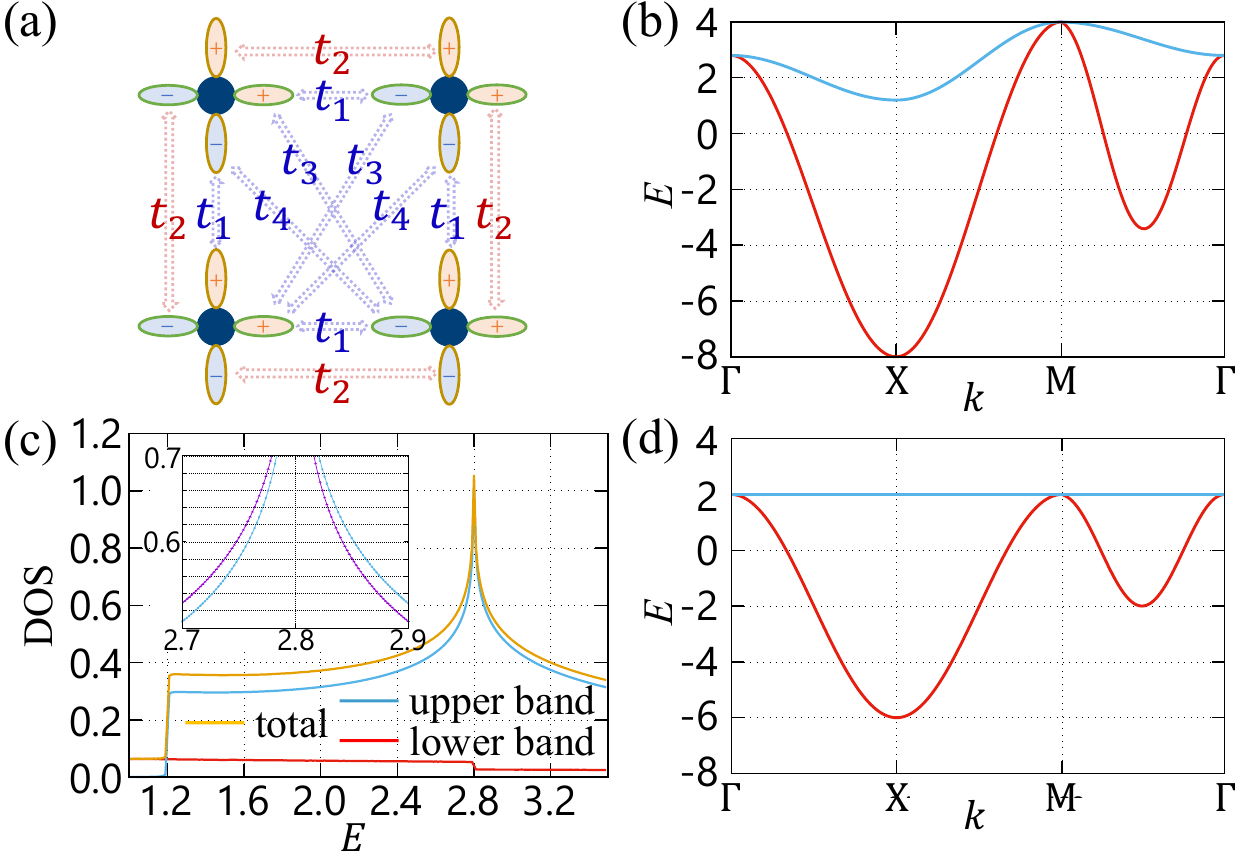}
  \centering
  \caption{(a) The hopping integrals in the 2D $t_{2g}$-orbital model. 
  (b) Band dispersion and (c) DOS for $(t_1, t_2, t_3, t_4) = (-1.0, 1.3, -0.85, -0.85)$. The red and blue lines in Fig.~\ref{fig:Raghu_hk}(c) show the DOS of the lower and upper bands, $D_-(E)$ and $D_+(E)$, respectively. The yellow line shows the total DOS. The inset compares $D_+(E)$ (blue) with $D_+(5.6-E)$ (purple), showing the asymmetry around the singular saddle point $E=2.8$. 
  (d) Band dispersion for the flat-band parameter, $t_1 =-t_2 = 2t_3=2t_4 = -1$.} \label{fig:Raghu_hk}
\end{figure}

The lattice model is given by $H(\bm k) = b_0(\bm k)\sigma_0 + \bm b(\bm k)\cdot\bm \sigma$ with $b_0(\bm k) = -(t_1+t_2)(\cos k_x+\cos k_y) -4t_3\cos k_x\cos k_y$, $b_{z}(\bm k) = -(t_1-t_2)(\cos k_x-\cos k_y)$, and $b_{x}(\bm k) = -4t_4\sin k_x \sin k_y$,
whose hopping integrals are schematically shown in Fig.~\ref{fig:Raghu_hk}(a).
The energy dispersion is given by $E_\pm(\bm k) = b_0(\bm k)\pm\vert\bm b(\bm k)\vert$.
Since the $d_{xz}$ and $d_{yz}$ orbitals belong to a 2D irreducible representation of the point group containing $C_{4}$ rotation symmetry, the model accommodates the band touching at the $\Gamma$ and $M$ points in the Brillouin zone.

For a certain parameter set $(t_1, t_2, t_3, t_4) = (-1.0, 1.3, -0.85, -0.85)$, this model is known as Raghu's model which can reproduce the Fermi surfaces of iron-based superconductors~\cite{Raghu2008}. 
As shown in Fig.~\ref{fig:Raghu_hk}(b), a singular saddle point appears at the $\Gamma$ point, where the upper band hosts a singular saddle point although the lower band shows a nearly parabolic dispersion. In contrast, the $M$ point accommodates a parabolic band touching. 
The DOS evaluated by $D(E) = \sum_{n = \pm}\int_{\rm BZ} \frac{d\bm k}{(2\pi)^2}\delta/\pi[(E_n(\bm k)-E)^2 + \delta^2]$ with $\delta = 0.001$ is shown in Fig.~\ref{fig:Raghu_hk}(c). Consistent with the analysis of the $kp$ Hamiltonian (Table~\ref{table:DOS}), the DOS of the singular band, $D_+(E)$ (blue line), shows a divergent behavior at the singular saddle point $E = 2.8$, while the DOS of the nearly parabolic band, $D_-(E)$ (red line), is almost constant with a discontinuous jump at $E = 2.8$. 
The asymmetric term $\theta(E-2.8)D_{\rm asym}$ expected from Table~\ref{table:DOS} also appears, as shown in the inset of Fig.~\ref{fig:Raghu_hk}(c). 
Thus, all features of the singular saddle point are reproduced in the $t_{2g}$-orbital model. 

The $t_{2g}$-orbital model can also model a flat-band system; the upper band becomes completely flat when we set $t_1 =-t_2 = 2t_3=2t_4$, as shown in Fig.~\ref{fig:Raghu_hk}(d). This is a kind of the singular flat band~\cite{Bergman2008-nc,Rhim2019}.
In this case, the flat band satisfies the Mielke's theorem~\cite{Mielke1999,Tasaki2020} for flat-band ferromagnetism, thereby ensuring that a half-filled flat band has a unique ferromagnetic ground state~\cite{Appendix}. Consequently, the 2D $t_{2g}$-orbital model serves as a theoretical framework for elucidating the link between QGFM and flat-band ferromagnetism.

Now we examine the criterion for QGFM, namely, the sign of $\chi_{\rm c}^{xx}$, based on the $kp$ Hamiltonian derived from the $t_{2g}$-orbital model. In the following, we focus on the band-touching $\Gamma$ point and set the chemical potential $\mu = b_0(0)$.
We assume $t_2 > -t_1 \geq 0$ and $-t_3 = -t_4 = t_{xy} >0$, for which the band dispersion is similar to Fig.~\ref{fig:Raghu_hk}(b). Since the unit of energy is set by $t_1 =-1$, the two parameters $a = t_2 - \vert t_1\vert$ and $t_{xy}$ determine the model.
In this setup, the upper band has a singular saddle point when $t_{xy} > a/2+1/2$ is satisfied, while the lower band has it when $t_{xy} < a/8$.

\begin{figure}[tbp]
  \includegraphics[width=1.0\linewidth]{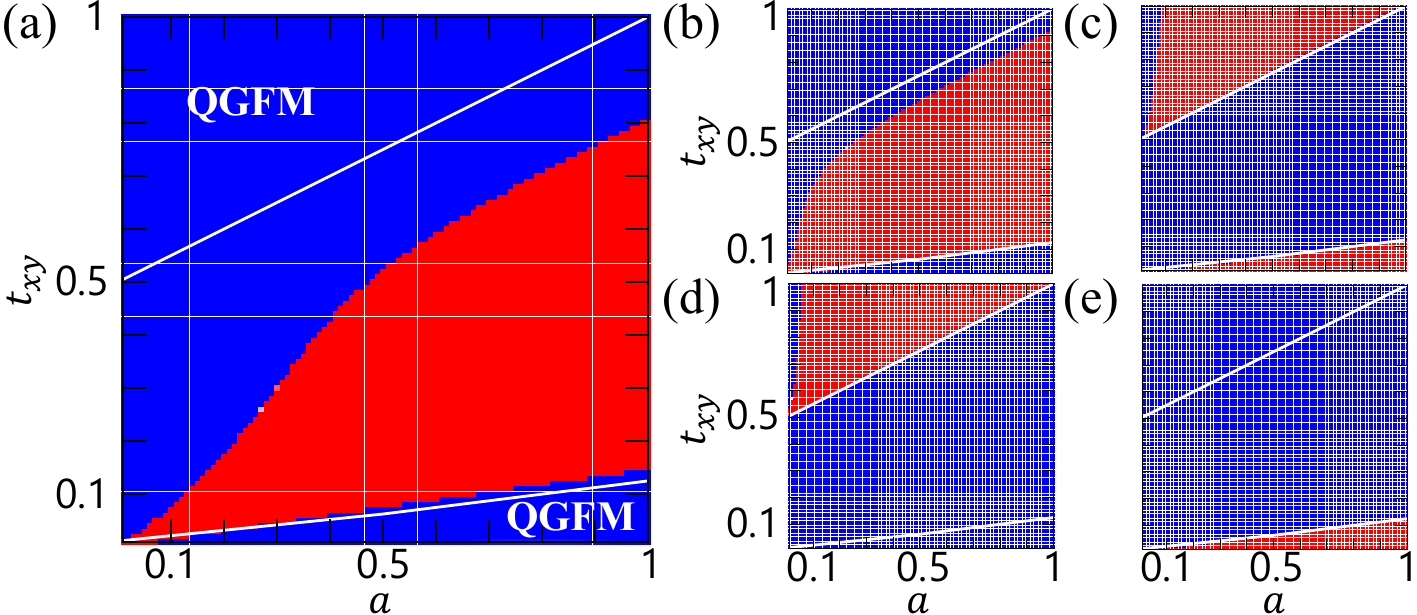}
  \centering
  \caption{The magnetic phase diagram of the 2D $t_{2g}$-orbital model at $T = 0.005$. The color shows the sign of $\chi_{\rm c}^{xx}$ and its components: (a) $\chi_{\rm c}^{xx}$, (b) $\chi_{\rm geom}^{xx}$, and (c) $\chi_{\rm mass}^{xx}$. $\chi_{\rm mass}^{xx}$ is the sum of contributions from the upper and lower bands, which are shown in (d) and (e), respectively.  
  These quantities are negative (positive) in the blue (red) region. 
  The white lines show $t_{xy} = a/2+1/2$ and $t_{xy} = a/8$. 
  \label{fig:Raghu_pd}}
\end{figure}

In Fig.~\ref{fig:Raghu_pd}, we show the magnetic phase diagram, where the red and blue colors represent the positive and negative signs of $\chi_{\rm c}^{xx}$ [Fig.~\ref{fig:Raghu_pd}(a)] and its components such as $\chi_{\rm geom}^{xx}$ and $\chi_{\rm mass}^{xx}$. 
The upper and lower white lines represent the lines $t_{xy} = a/2+1/2$ and $t_{xy} = a/8$, respectively. Therefore, a singular saddle point appears outside the two white lines.
Consistent with the analytic discussions above, the quantum geometric term $\chi_{\rm geom}^{xx}$ [Fig.~\ref{fig:Raghu_pd}(b)] is mostly negative and favors ferromagnetic correlation when the singular saddle point appears. Otherwise, the quantum geometric term mostly favors antiferromagnetic correlation. 

The effective mass term $\chi_{\rm mass}^{xx}$ [Fig.~\ref{fig:Raghu_pd}(c)] competes with the quantum geometric term in the large parameter range.
As we show in Figs.~\ref{fig:Raghu_pd}(d) and \ref{fig:Raghu_pd}(e), the band with a saddle point gives a positive contribution and makes $\chi_{\rm mass}^{xx}$ positive. 
In the absence of the saddle point, the effective mass term is negative and favors ferromagnetic correlation, as in the case of the parabolic band~\cite{Kitamura2024}. 

\begin{figure}[t]
  \includegraphics[width=1.0\linewidth]{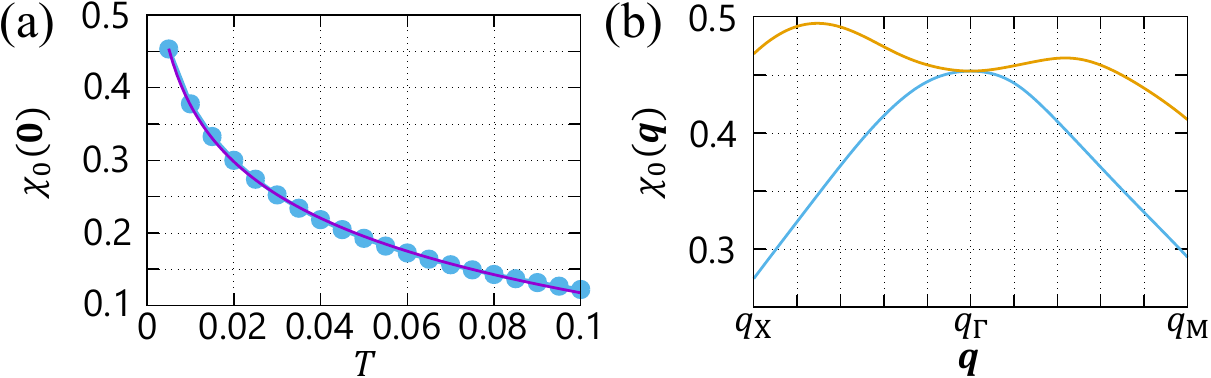}
  \centering
  \caption{
  (a) The temperature dependence of ferromagnetic spin susceptibility $\chi_0(0)$ for $(t_1, t_2, t_3, t_4) = (-1.0, 1.3, -0.85, -0.85)$. 
  The purple line is the fitting curve, $-0.112297\log T \TK{-}0.141179$. (b) Momentum dependence of spin susceptibility $\chi_{0}(\bm q)$ (blue line) and that obtained for the geometrically trivial model $H_{\rm kp}(\bm p) = {\rm diag}[\epsilon_+(\bm p),\epsilon_-(\bm p)]$ (orange line) at $T=0.005$. The susceptibilities are shown on the symmetry lines from $\bm q_{\Gamma} = (0,0)$ to
  $\bm q_{X} = (\pi/8,0)$ and to $\bm q_{M} = (\pi/8,\pi/8)/\sqrt{2}$. 
  }
  \label{fig:Raghu_chi}
\end{figure}

Comparing the total $\chi_{\rm c}^{xx}$ in Fig.~\ref{fig:Raghu_pd}(a) with the quantum geometric term in Fig.~\ref{fig:Raghu_pd}(b), we find that the quantum geometry governs magnetism in almost all cases. In particular, in the parameter region with a singular saddle point, ferromagnetic fluctuation almost always appears mainly due to quantum geometry. 
Consistent with the analysis of $\chi_{\rm c}^{xx}$, we confirm the existence of the peak in spin susceptibility at $\bm q = 0$ [blue line in Fig.~\ref{fig:Raghu_chi}(b)]. 
In contrast, when we neglect quantum geometry and calculate the model $H_{\rm kp}(\bm p) = {\rm diag}[\epsilon_+(\bm p),\epsilon_-(\bm p)]$, $\chi_{0}(\bm q)$ has peaks away from $\bm q =0$ [orange line in Fig.~\ref{fig:Raghu_chi}(b)] indicating the antiferromagnetic correlation.
In addition, the bare spin susceptibility for ferromagnetism $\chi_0(0)$ shows a logarithmic divergence due to the singular DOS, as confirmed by the fitting in Fig.~\ref{fig:Raghu_chi}(a). 
Therefore, the ferromagnetic spin susceptibility is divergent at zero temperature, and switching on the Coulomb interaction leads to the ferromagnetic order. 
Thus, we conclude that QGFM appears ubiquitously when the singular saddle point lies on the Fermi energy.

\begin{figure}[tbp]
  \includegraphics[width=1.0\linewidth]{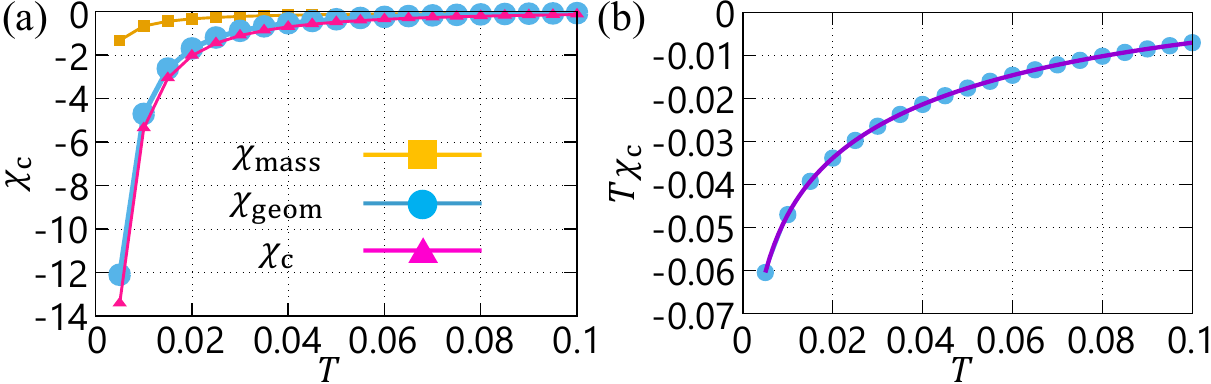}
  \centering
  \caption{
  (a) The curvature of spin susceptibility $\chi_{\rm c}^{xx}$ (pink line) for $a=0$ and $t_{xy}=0.5$, in which a singular flat band appears [Fig.~\ref{fig:Raghu_hk}(d)]. The orange and blue lines show $\chi_{\rm mass}^{xx}$ and $\chi_{\rm geom}^{xx}$, respectively. (b) Fitting of $T\chi_{\rm geom}$ (blue dots) by the function $0.0198944\log T + 0.0452354 -0.0645027T$. \label{fig:Raghu_chi_flat}}
\end{figure}

\textit{From QGFM to flat-band ferromagnetism---}The magnetic phase diagram in Fig.~\ref{fig:Raghu_pd}(a) contains the flat-band parameter $a=0$ and $t_{xy}=0.5$ [Fig.~\ref{fig:Raghu_hk}(d)], where we exactly prove flat-band ferromagnetism in the Appendix~\cite{Appendix}. Therefore,
flat-band ferromagnetism can be viewed as an extreme case of QGFM. 
Let us characterize flat-band ferromagnetism from the viewpoint of QGFM.
In the singular flat-band system, $I(\theta)$ in Eq.~\eqref{eq:geom_theta} exhibits a logarithmic divergence for all $\theta$ according to the $T\to0$ formula Eq.~\eqref{eq:zero_temp_I}, which is replaced by $\log T$ behavior at finite temperatures. Thus, the quantum geometric term in $\chi_{\rm c}^{ij}$ exhibits a negative divergence of $\frac{1}{T}\log T$~\cite{Appendix}. 
This is consistent with our numerical results in Fig.~\ref{fig:Raghu_chi_flat}, which show that $T\chi_{\rm geom}^{xx}$ is well fitted by $0.0198944\log T + 0.0452354 -0.0645027T$.
Due to this contribution arising from the divergent quantum metric, the singular flat-band systems show ferromagnetic correlation at low temperatures, and the infinite DOS ensures ferromagnetic order according to the Stoner theory. Thus, a combination of quantum geometry and divergent DOS provides a concise view of flat-band ferromagnetism, the exactness of which in turn implies the stability of ferromagnetism in a wide range of systems accommodating singular saddle points.

In general, the overlap of Wannier functions is known to be essential for the uniqueness of flat-band ferromagnetism~\cite{Tasaki2020}. Considering that the gauge-invariant part of the spread of a Wannier function is given by the quantum metric~\cite{Marzari1997,Vanderbilt2018}, the understanding based on QGFM is closely related to that based on the overlap of Wannier functions. 
Thus, ferromagnetism in the singular flat band is naturally taken over to QGFM in the singular saddle point band through the quantum metric, which, in turn, implies that QGFM can survive beyond the Stoner theory.
Although it is difficult to achieve the flat-band parameter experimentally, the ferromagnetism can be robust due to a singular saddle point when the flat band becomes dispersive. This broadens the scope of materials capable of ferromagnetism that share its origin with flat-band ferromagnetism.

\textit{Discussion---}Recently, itinerant ferromagnetism was newly reported in lithium-intercalated FeSe~\cite{Hu2025-dt}. Unfortunately, detailed band calculation results from density functional theory (DFT) have not been provided for this compound. However,  strong quantum geometric effects in iron-based superconductors, including FeSe~\cite{Kitamura2021-jk,Kitamura2022-wi}, have been revealed by the DFT study. Thus, controlling the band structure of iron-based superconductors may realize QGFM by a singular saddle point, and lithium-intercalated FeSe is a candidate system.

Although we have assumed $C_4$ symmetry in this Letter, other symmetries such as $C_3$ symmetry can protect band touching, and the singular saddle point can appear. 
In fact, we can show that the kagome lattice model with extra hopping hosts a singular saddle point and exhibits ferromagnetism~\cite{Supple}.
Our proposal for QGFM by the singular saddle point can also be applied to such systems and would be a guiding principle for the exploration of 2D ferromagnetic materials.

\begin{acknowledgments}
We are grateful to Y. Sigedomi, M. Tezuka, T. Miki, T. Nomoto, and R. Arita, for fruitful discussions.
This work was supported by JSPS KAKENHI (Grant Nos. JP21K13880, JP22H01181, JP22H04476, JP22H04933, JP22J22520, JP23K17353, JP23K22452, JP23KJ0783, JP24K21530, JP24H00007, JP25H01249, \TK{JP24H01662, JP23KK0248}).
\end{acknowledgments}

\clearpage

\section{EndMatter}    

\textit{DOS in models with a singular saddle point---}We show the derivation of the DOS for the energy bands $\epsilon_{\pm}^{\rm eff}(\bm p)$.
Although we consider the case of $1/ M_{xy}>1/\vert M\vert > 1/M_{xx}$ and $2/M_{xx} > 1/M_{xy}-1/\vert M \vert$, 
DOS for other parameters can be derived in the same manner.
Because of $C_{4}$ symmetry, the DOS is given by
$
  D_\pm(\varepsilon) = 4\int_{S}[{d\bm p}/{(2\pi)^2}]\delta(\epsilon_\pm(\bm p)-\varepsilon)
$.
where $S$ is the area satisfying $p_x^2 + p_y^2 \leq p_c^2$ and $p_x,p_y>0$.
The coordinates 
$p_{1(2)} = (p_x \pm p_y)/\sqrt{2}$ are useful for the calculation. In these coordinates, $S$ satisfies $p_1^2 + p_2^2 \leq p_c^2$ and $p_1 \geq \vert p_2\vert$ and the energy dispersion is rewritten by $\epsilon_{\pm}(\bm p) = \pm{p_{1}^2}/{2m_{1}^\pm}+ {p_2^2}/{2m_2^\pm}$ with $1/m_{1(2)}^{+}=\vert1/m^+ \pm 1/m_{xy}\vert$ and $1/m_{1(2)}^{-} = \vert1/m^- \mp 1/m_{xy}\vert$.

First, we evaluate $D_-(\varepsilon)$. For $\varepsilon > 0$, by the variable transformation, $p_1 = p[m_1^-]^{1/2}\sinh(\theta)$ and $p_2 = p[m_2^-]^{1/2}\cosh(\theta)$, we get
$
  D(\varepsilon) 
  =\int d\theta D^0_-
$
with 
$
D^0_\pm = (m_1^\pm m_2^\pm)^{\frac{1}{2}}/\pi^2
$.
The integral area of $\theta$ is determined by the condition 
$
( m_1^-/m_2^--1) ^{-\frac{1}{2}}\leq \vert\sinh\theta\vert
\leq
(p_c^2/2\varepsilon - m_2^-)^{\frac{1}{2}}
/(m_1^- + m_2^-)^{\frac{1}{2}}.
$
The lower and upper bounds are owing to the conditions $p_1 \geq \vert p_2\vert$ and $p_1^2 + p_2^2 \leq p_c^2$, respectively.
Leaving $\mathcal{O}(\vert\varepsilon\vert\cdot2m^-_{1(2)}/p_c^2)$, we obtain $D_-(\varepsilon ) = D_{\rm log}(\varepsilon) - D_{\rm asym} $, with
$    
    D_{\rm log}(\varepsilon)=
    -D^0_-\ln\vert\varepsilon\vert 
    + D^0_-\ln [2p_c^2/(m_1^- + m_2^-)]
$
and
$    
    D_{\rm asym}= 2D^0_-
    \ln(\vert1-m_2^-/m_1^-\vert ^{-\frac{1}{2}} + \vert1-m_1^-/m_2^-\vert ^{-\frac{1}{2}}).
$ 
For $\varepsilon<0$, we use the variable transformation $p_1 = p[m_1^-]^{1/2}\cosh\theta$ and $p_2 =p[m_2^-]^{1/2} \sinh\theta$. In this case, since $p_1 > \vert p_2\vert$ is always satisfied, the lower bound of $\sinh\theta$ is zero, and we get $D_-(\varepsilon ) = D_{\rm log}(\varepsilon)$.
Second, for $D_{+}(\varepsilon)$, using the polar coordinate, we can easily derive $D_{+}(\varepsilon) = \theta(\varepsilon)D_{\rm pb}$ with $ D_{\rm pb}= 2\arcsin  [m_1
/(m_1 + m_2)]^{1/2} D^0_+$.

\textit{Criterion for ferromagnetic fluctuation---}The spin susceptibility of $H_{\rm kp}(\bm p)$ is given by
\begin{align}
  \chi_0(\bm q) &= \sum_{m,m^\prime=\pm}\int_{\vert \bm p\vert\leq p_{\rm c}}
  \dfrac{d\bm p}{(2\pi)^2}
  \dfrac{f(\epsilon_m(\bm p+\bm q))-f(\epsilon_{m^\prime}(\bm p))}{\epsilon_{m^\prime}(\bm p)-\epsilon_{m}(\bm p+\bm q)}\notag\\
  &\times{\rm tr}[P_{m}(\bm p+\bm q)P_{m^\prime}(\bm p)],
\end{align}
with $P_\pm(\bm p) = [\vert\bm h(\bm p)\vert\pm \TK{\bm h(\bm p)\cdot \bm \sigma}]/2\vert\bm h(\bm p)\vert$. Here, ${\rm tr}$ represents the trace of two-by-two matrices.
The curvature of spin susceptibility $\chi_{\rm c} = \chi_{\rm geom} + \chi_{\rm mass}$ is given by,
\begin{eqnarray}
    \chi_{\rm geom}^{ij} &=& \sum_{m=\pm}\int_{\vert\bm p\vert \leq p_c}\dfrac{d\bm p}{(2\pi)^d}
    \dfrac{\partial_{p_i}\hat{\bm h}(\bm p)\cdot\partial_{p_j}\hat{\bm h}(\bm p)}{2}\notag\\
    &\times&\left[
    f^\prime(\epsilon_m(\bm p))    
    -s_m{f(\epsilon_m(\bm p))}/{\vert \bm h(\bm p)\vert}
    \right],~\label{eq:geomeric_term}\\
    \chi_{\rm mass}^{ij} &=& -\sum_{m = \pm}\int_{\vert\bm p\vert \leq p_c}\dfrac{d\bm p}{(2\pi)^d}
    \partial_{p_i}\partial_{p_j}\epsilon_m(\bm p)\dfrac{f^{(2)}(\epsilon_m(\bm p))}{6},\notag\\
\end{eqnarray}
with the Fermi distribution function $f(\epsilon)$ and $s_\pm = \pm$. Note that $\partial_{p_i}\hat{\bm h}(\bm p)\cdot\partial_{p_j}\hat{\bm h}(\bm p)/2$ represents the quantum metric of two-band systems. 

\textit{Effective mass term---}We show the derivation of the effective mass term $\chi_{\rm mass}^{ii}$ for the energy dispersion, $\epsilon_{\pm}^{\rm eff}(\bm p) = \frac{p_{x}^2+p_{y}^2}{2m^{\pm}} \pm \frac{\vert p_xp_y\vert} {m_{xy}}$ with saddle-point structure.
By ignoring the discontinuity of $\partial_x^2 \epsilon_{\pm}^{\rm eff}(\bm p)$ at $p_x = 0$, the effective mass contributed from the band with a saddle point is given by,
\begin{align}
    \chi_{\rm mass:sp}^{ii}
    &=-\int_{-\infty}^{\infty} d\varepsilon D_\pm(\varepsilon)\dfrac{f^{(2)}(\varepsilon)}{6m^\pm}\notag\\
    &=\int_{0}^{\infty} d\varepsilon D_{\rm asym}\dfrac{f^{(2)}(\varepsilon)}{6m^\pm}=\dfrac{D_{\rm asym}}{24T\vert m^{\pm}\vert},
\end{align}
where we used the fact that $D_{\rm log}(\varepsilon)f^{(2)}(\varepsilon)$ is an odd function. 
The other contribution $\chi_{\rm mass:pb}^{ii}$ is calculated in the same way.

\textit{Quantum geometric term---}We start from Eq.~\eqref{eq:geomeric_term} and assume that the order of $\epsilon_n(\bm p)$ is $\vert \bm p\vert^n$.
For the calculation, we use the polar coordinates $(p_{x}, p_{y}) = p(\cos\theta, \sin\theta)$. 
The quantum metric of two-band systems and energy dispersion are described by the polar coordinates as, 
$\partial_{p_i}\hat{\bm h}(\TK{\bm p})\cdot\partial_{p_j}\hat{\bm h}(\TK{\bm p})= g^{ij}(\theta)/p^{2}$  and $\epsilon_\pm(\bm p) = p^n\epsilon_\pm(\theta)$.
Therefore, $\chi_{\rm geom}^{ij}$ is rewritten by,
\begin{align}
    \chi_{\rm geom}^{ij} &= \dfrac{1}{4\pi^2}\int_{0}^{2\pi}d\theta\int_{0}^{p_{\rm c}}dp
    \dfrac{g^{ij}(\theta)}{2}\notag\\
    &\times\sum_{m=\pm}\left(
    \dfrac{f^\prime(p^n\epsilon_m(\theta))}{p}
    -s_m\dfrac{f(p^n\epsilon_m(\theta))}{p^{n+1}\vert \bm h(\theta)\vert}
    \right)\notag\\
    &=\dfrac{1}{4\pi^2}\sum_{m=\pm}s_m\int_{0}^{2\pi}d\theta
    \dfrac{g^{ij}(\theta)}{2}\left\{
    \left[
    \dfrac{f(p^n\epsilon_m(\theta))}{np^{n}\vert \bm h(\theta)\vert}
    \right]_0^{p_{\rm c}}\right.
    \notag\\
    &-\left.\int_{0}^{p_{\rm c}}dp
    \dfrac{h_0(\theta)}{\vert \bm h(\theta)\vert}
    \dfrac{f^\prime(p^n\epsilon_m(\theta))}{p}
    \right\}\notag\\
    &\approx\dfrac{1}{4\pi^2}\int_{0}^{2\pi}d\theta
    \dfrac{g^{ij}(\theta)}{2nT}\left\{
    \dfrac{1}{2} + I(\theta)\right\},\\
    I(\theta)&=\int_{0}^{\frac{p_{\rm c}^n}{T}}dp
    \dfrac{h_0(\theta)}{\vert \bm h(\theta)\vert}
    \dfrac{l^\prime(p\epsilon_-(\theta))-l^\prime(p\epsilon_+(\theta))}{p},\label{eq:appendix_geom}
\end{align}
where $l(x)$ is defined by $l(x) = f(Tx)$. In the final equation, we ignore the term $\mathcal{O}(T^0)$.

Then, we derive the low-temperature formula of $I(\theta)$. In the following equations, we omit the variable $\theta$ for simplicity.
First, in the case of $\vert h_0\vert \neq \vert\bm h\vert$, we get
\begin{align}
    &\dfrac{\vert \bm h\vert}{h_0}I=-
    \left[\int_{0}^{\delta} + \int_{\delta}^{\frac{p_{\rm c}^n}{T}}\right]dp\sum_{m = \pm}\dfrac{s_ml^\prime(p\epsilon_m)
	}{p}
    \notag\\
    &=
    \left[\int_{\delta\vert \epsilon_-\vert}^{\frac{p_{\rm c}^\TK{n}\vert \epsilon_-\vert}{T}}
	-\int_{\delta\vert\epsilon_+\vert}^{\frac{p_{\rm c}^\TK{n}\vert\epsilon_+\vert}{T}}
    \right]dp\dfrac{l^\prime(p)
    }{p}
    -\sum_{m = \pm}\int_{0}^{\delta}dp\dfrac{s_ml^\prime(p\epsilon_m)
	}{p}
    \notag\\
    &\xrightarrow{T=0}
    \int_{\delta\vert \epsilon_-\vert}^{\delta\vert\epsilon_+\vert}dp
    \dfrac{l^\prime(p)
    }{p}
    -\sum_{m=\pm}s_m\int_{0}^{\delta}dp\dfrac{l^\prime(p\epsilon_m)}{p},~\label{eq:appendix_I_zero}
\end{align}
where $\delta \cdot {\rm max}[\vert\epsilon_\pm\vert]\ll 1$ is satisfied.
Therefore, we can use the Taylor expansion of $l^\prime(x)$ and integrate the first and second terms of Eq.~\eqref{eq:appendix_I_zero} as,
\begin{align}
    &\int_{\delta\vert \epsilon_-\vert}^{\delta\vert\epsilon_+\vert}dp
    \dfrac{l^\prime(p)
    }{p} = -\dfrac{1}{4}\ln\left\vert\dfrac{\epsilon_+}{\epsilon_-}\right\vert\notag\\
    &\,\,\,\,\,\,\,\, +\sum_{x=1}^{\infty}\dfrac{\delta^{x}l^{(x+1)}(0)([\epsilon_+]^x-[\epsilon_-]^x)}{x\cdot(x
    !)},~\label{eq:appendix_I2}\\
    &-\sum_{m=\pm}s_m\int_{0}^{\delta}dp\dfrac{l^\prime(p\epsilon_m)}{p} =
    \notag\\
    &
    \,\,\,\,\,\,\,\, 
    \sum_{x=1}^{\infty}\dfrac{\delta^{x}l^{(x+1)}(0)([\epsilon_-]^x-[\epsilon_+]^x)}{x\cdot(x
    !)}.~\label{eq:appendix_I1}
\end{align}
By summing up Eqs.~\eqref{eq:appendix_I1} and ~\eqref{eq:appendix_I2} and inserting them into $I$, we get Eq.~\eqref{eq:zero_temp_I}.
Second, in the case {of}  $\vert h_0\vert = \vert\bm h\vert$,
$I$ is rewritten by,
\begin{align}
    &I 
    =\int_{0}^{p_{\rm c}^n/T}dp
	\left(-\dfrac{1}{4p}
    -\dfrac{l^\prime(2p\vert \bm h\vert)
	}{p}\right)\notag\\
    &=\left[\int_{0}^{\delta}
    +\int_{\delta}^{2\vert\bm h\vert p_{\rm c}^n/T}\right]dp
	\left(-\dfrac{1}{4p}
    -\dfrac{l^\prime(p)
	}{p}\right),
\end{align}
with $\delta \ll 1$.
The convergence of the first integral can be proven by using the Taylor expansion of $l^\prime(p)$. 
However, for the second integral, while $\TK{-}\int_\delta^{2\vert\bm h\vert p_{\rm c}^n/T}dp{l^\prime(p)}/{p}< \int_\delta^{2\vert\bm h\vert p_{\rm c}^n/T}dp{e^{-p}}/{p}$ converges, $\int_\delta^{2\vert\bm h\vert p_{\rm c}^n/T} dp/p$ shows the divergence in the low-temperature limit as $\log T$.

\begin{widetext}

\textit{Flat-band ferromagnetism---}Here, we sketch the proof of the flat-band ferromagnetism in the 2D $t_{2g}$-orbital model with the on-site Coulomb interaction $U$ for the parameter, $t_1 =-t_2 = 2t_3=2t_4$.
Hereafter, we focus on the Hilbert space for up-spin electrons, which is enough for the proof.
For simplicity, we assume that the total number of unit cells, $L^2 = N$, is even. 
In that case, the wave-number $\bm k = (2\pi r_x/L,2\pi r_y/L)$ with non-negative integers $0\leq r_{x}, r_y\leq L-1$ includes $\Gamma$ and $M$ points where the flat band degenerate with the other dispersive band.
Therefore, there are $N+2$ single-particle degenerate states.
For preparation, we introduce the compact localized state (CLS), which is the eigenstate of the flat band and given by,
$
    \ket{\alpha_{\bm r}} = (\hat{c}_{x,\bm r}^\dagger + \hat{c}_{x,\bm r+\hat{\bm x}}^\dagger - \hat{c}_{x,\bm r+\hat{\bm y}}^\dagger - \hat{c}_{x,\bm r+\hat{\bm x}+\hat{\bm y}}^\dagger
    - \hat{c}_{y,\bm r}^\dagger + \hat{c}_{y,\bm r+\hat{\bm x}}^\dagger - \hat{c}_{y,\bm r+\hat{\bm y}}^\dagger + \hat{c}_{y,\bm r+\hat{\bm x}+\hat{\bm y}}^\dagger)\ket{0},
$
where we define $\bm r = (r_x, r_y), \hat{\bm x} = (1,0)$ and $\hat{\bm y} = (0,1)$. $\hat{c}^\dagger_{x(y),\bm r}$ is the creation operator for the $d_{xz(yz)}$ orbital at $\bm r$ and
$\ket{0}$ is the vacuum state.
Since the flat band degenerates with the other band, the $N$ translation copies of CLS do not span the linearly independent basis of the flat band. Instead, $N-2$ translation copies of CLS and $4$ noncontractible loop state (NLS), which 
is extended in only one direction, spans linearly independent basis of the flat band~\cite{Bergman2008-nc,Rhim2019}.
The $4$ NLSs are given by,
$
  \ket{\gamma_{x,r_y}}= \sum_{r_x}\hat{c}^\dagger_{x,\bm r}\ket{0},
  \ket{\gamma_{y,r_x}}= \sum_{r_y}\hat{c}^\dagger_{y,\bm r}\ket{0},
  \ket{\mu_{y,r_y}}= \sum_{r_x}(-1)^{r_x}\hat{c}^\dagger_{y,\bm r}\ket{0},
  \ket{\mu_{x,r_x}}= \sum_{r_y}(-1)^{r_y}\hat{c}^\dagger_{x,\bm r}\ket{0}.
$

Then, for the proof, we follow the standard strategy~\cite{Mielke1999,Tasaki2020}:
In general, the quasi-local state (QLS), $\ket{\rho_{\tau}}$, which satisfies $\bra{\rho_\tau}\hat{P}\ket{\rho_{\tau^\prime}}\propto\delta_{\tau,\tau^\prime}$ with $\tau = (l,\bm r)\in \Lambda_\rho$ and the indices of orbitals $l$, spans linearly independent basis of flat band. Here, $\Lambda_\rho$ is the subset of orbitals with total number $N+2$, and $\hat{P}$ is the projection operator onto the orbitals of $\Lambda_\rho$. If, for any $\tau,\tau^\prime\in\Lambda_\rho$, there is the sequence $\tau_0,\ldots,\tau_{n}$ such that $\tau_0 = \tau,\tau_{n} = \tau^\prime$, and $\braket{\rho_{\tau_{j-1}} \vert\rho_{\tau_{j}}}\neq 0$ for all $j = 1,\ldots,n$, namely the connectivity condition, the half-filled flat band has a unique ferromagnetic ground state.
In the 2D $t_{2g}$-orbital model, $\Lambda_{\rho}$ is divided into two subsets $\Lambda_{x}$ and $\Lambda_{y}$ which is constructed by $d_{xz}$ and $d_{yz}$ orbitals, respectively, and the total number of each subset is $N/2+1$.
By the combination between CLS and NLS, for $\tau_x \in\Lambda_x$ and $\tau_y\in\Lambda_y$ with $r_x \in \{0,2,\dots,L-2\}$, QLS is given by~\cite{Supple} 
\begin{align}
    &\ket{\rho_{\tau_x}} = \left\{
    \begin{array}{ccc}
    \ket{\phi_{\bm r}} =
    \sum_{r_y^\prime = 0}^{r_y\TK{-1}}\ket{+_{(r_x,r_y^\prime)}}
    +\ket{\gamma_{y,r_x-1}}-\ket{\gamma_{y,r_x+1}}
    - \ket{\mu_{x,r_x-1}} - \ket{\mu_{x,r_x+1}} &
    {\rm for} &
    r_x\neq r_x^*\ \cap\ r_y\neq0\\
    \ket{\mu_{x,r_x+1}} &
    {\rm for} &
     r_y=0\\
    \ket{\phi_{\bm r}} + 2\sum_{n = 1}^{L/2}\ket{\mu_{x,2n-1}}
    -2\ket{\gamma_{x,0}} &
    {\rm for} &
    r_x = r_x^*\ \cap\ r_y\neq0\\
    \ket{\gamma_{x,r_y}} - \sum_{n = 1}^{L/2}\ket{\mu_{x,2n-1}}&
    {\rm for} &
    r_x = r_x^*\ \cap\ r_y=0\\
    \end{array},
    \right.\\
    &\ket{\rho_{\tau_y}} = \left\{
    \begin{array}{ccc}
    \ket{\psi_{\bm r}} = \sum_{r_y^\prime = 0}^{r_y\TK{-1}}(-1)^{r_y^\prime}\ket{-_{(r_x,r_y^\prime)}}
    +\ket{\gamma_{y,r_x-1}}+\ket{\gamma_{y,r_x+1}}
    - \ket{\mu_{x,r_x-1}} + \ket{\mu_{x,r_x+1}} &
    {\rm for} &
    r_x\neq r_x^*\ \cap\ r_y\neq0\\
    \ket{\gamma_{y,r_x+1}} &
    {\rm for} &
     r_y=0\\
    \ket{\psi_{\bm r}} \TK{-} 2\sum_{n = \TK{1}}^{L/2}\ket{\gamma_{\TK{y},2n-1}}
    -2\ket{\mu_{y,0}} &
    {\rm for} &
    r_x = r_x^*\ \cap\ r_y\neq0\\
    \ket{\mu_{y,r_y}} + \sum_{n = \TK{1}}^{L/2}\ket{\gamma_{y,2n-1}}&
    {\rm for} &
    r_x = r_x^*\ \cap\ r_y=0\\
    \end{array},
    \right. 
\end{align}
with $\ket{\pm_{\bm r}} = \pm\ket{\alpha_{\bm r}}+\ket{\alpha_{\bm r-\hat{\bm x}}}$ and arbitaly even $r_x^*$.
We can check the connectivity condition for each subset $\Lambda_{x}$ and $\Lambda_{y}$. In addition, because of $\braket{\rho_{d_{yz},(r_x,\TK{1})}\vert\rho_{d_{xz},(r_x\TK{+2},\TK{2})}}\neq0$, the above QLS satisfies the connectivity conditions whole in $\Lambda_{\rho}$, and thereby, a half-filled flat band has a unique ferromagnetic ground state.

\end{widetext}

\EnableTOC
\onecolumngrid
\clearpage
\setcounter{section}{0}
\setcounter{subsection}{0}
\setcounter{equation}{0}
\setcounter{figure}{0}
\setcounter{table}{0}
\renewcommand{\thesection}{S\arabic{section}}
\renewcommand{\theequation}{S\arabic{equation}}
\renewcommand{\thefigure}{S\arabic{figure}}
\renewcommand{\thetable}{S\arabic{table}}
\renewcommand{\theHsection}{supp.\arabic{section}}
\renewcommand{\theHequation}{supp.\arabic{equation}}
\renewcommand{\theHfigure}{supp.\arabic{figure}}
\renewcommand{\theHtable}{supp.\arabic{table}}
\setcounter{secnumdepth}{3}
\renewcommand{\LastBibLabel}{LastSuppBibItem}
\makeatletter
\let\@FMN@list\@empty
\makeatother
\newcommand*\wien{\textsc{wien}2k\xspace}

\begin{center}
 {\large \textmd{Supplemental Materials:} \\[0.3em]
 {\bfseries Quantum geometric ferromagnetism by singular saddle point}}
\end{center}


\section{Singular saddle point in kagome lattice model}
In this section, we show that the extended kagome lattice model hosts a singular saddle point protected by $C_{3}$ symmetry at the $\Gamma$ point.
The noninteracting Hamiltonian of the conventional kagome lattice model is given only by the nearest-neighbor (NN) hoppings as
\begin{align}
H_{\rm K}(\bm{k})
&=
\begin{pmatrix}
0 & e_{\rm AB}(\bm{k}) & e_{\rm AC}(\bm{k})\\
[e_{\rm AB}(\bm{k})]^{*} & 0 & e_{\rm BC}(\bm{k})\\
[e_{\rm AC}(\bm{k})]^{*} & [e_{\rm BC}(\bm{k})]^{*} & 0
\end{pmatrix},
\end{align}
where we define
\begin{align}
e_{\rm AB}(\bm{k})&=t\left(1+e^{-i\bm{k}\cdot\bm{\delta}_1}\right),\\ 
e_{\rm BC}(\bm{k})&=t\left(1+e^{-i\bm{k}\cdot\bm{\delta}_2}\right),\\ 
e_{\rm AC}(\bm{k})&=t\left(1+e^{i\bm{k}\cdot\bm{\delta}_3}\right),
\end{align}
with NN bond vectors $\bm{\delta}_1=(1,0),
\bm{\delta}_2=(-\frac12,\frac{\sqrt3}{2}),$ and $\bm{\delta}_3=(-\frac12,-\frac{\sqrt3}{2})$.
This model hosts a singular flat band~\cite{SBergman2008-nc,SRhim2019-km} whose band touching at the $\Gamma$ point is protected by the $C_{3}$ symmetry.

\begin{figure}[bp]
  \includegraphics[width=0.5\linewidth]{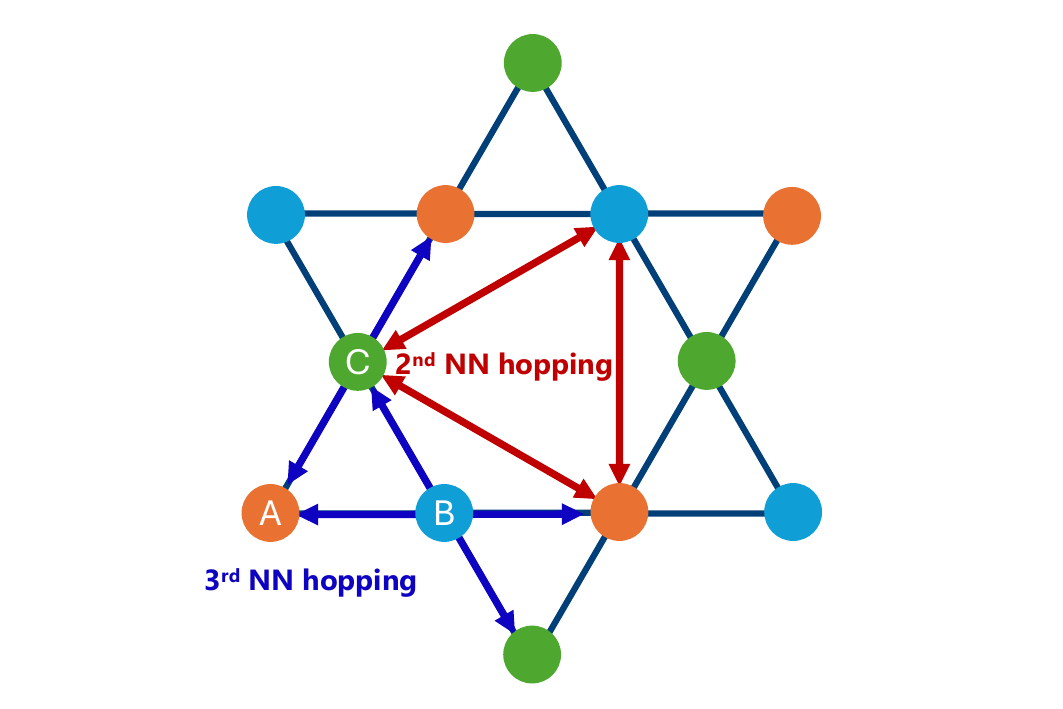}
  \centering
  \caption{Kagome lattice with the second- and third-NN hoppings. The orange, blue, and green-filled circles represent different sublattices.
      The red and blue arrows indicate second- and third-NN hoppings, respectively.
  }
  \label{fig:kaome_lattice}
\end{figure}

By adding the second- and third-NN hopping terms, the extended kagome lattice model is given by
\begin{align}
H_{\rm exK}(\bm{k})
&=
\begin{pmatrix}
e_{\rm AA}(\bm{k}) &
e_{\rm AB}(\bm{k})+e'_{\rm AB}(\bm{k}) &
e_{\rm AC}(\bm{k})+e'_{\rm AC}(\bm{k})\\
\bigl[e_{\rm AB}(\bm{k})+e'_{\rm AB}(\bm{k})\bigr]^{*} &
e_{\rm BB}(\bm{k}) &
e_{\rm BC}(\bm{k})+e'_{\rm BC}(\bm{k})\\
\bigl[e_{\rm AC}(\bm{k})+e'_{\rm AC}(\bm{k})\bigr]^{*} &
\bigl[e_{\rm BC}(\bm{k})+e'_{\rm BC}(\bm{k})\bigr]^{*} &
e_{\rm CC}(\bm{k})
\end{pmatrix},
\end{align}
where we define
\begin{align}
e'_{\rm AB}(\bm{k})
&= t^\prime\left(e^{i\bm{k}\cdot\boldsymbol{\delta}_2}+e^{i\bm{k}\cdot\boldsymbol{\delta}_3}\right),\\
e'_{\rm AC}(\bm{k})
&= t^\prime\left(e^{-i\bm{k}\cdot\boldsymbol{\delta}_2}+e^{-i\bm{k}\cdot\boldsymbol{\delta}_1}\right),\\
e'_{\rm BC}(\bm{k})
&= t^\prime\left(e^{i\bm{k}\cdot\boldsymbol{\delta}_1}+e^{i\bm{k}\cdot\boldsymbol{\delta}_3}\right),\\
e_{\rm AA}(\bm{k})
&= 2t^{\prime\prime}\left[\cos\!\bigl(\bm{k}\cdot\boldsymbol{\delta}_1\bigr)
          +\cos\!\bigl(\bm{k}\cdot\boldsymbol{\delta}_3\bigr)\right],\\
e_{\rm BB}(\bm{k})
&= 2t^{\prime\prime}\left[\cos\!\bigl(\bm{k}\cdot\boldsymbol{\delta}_1\bigr)
          +\cos\!\bigl(\bm{k}\cdot\boldsymbol{\delta}_2\bigr)\right],\\
e_{\rm CC}(\bm{k})
&= 2t^{\prime\prime}\left[\cos\!\bigl(\bm{k}\cdot\boldsymbol{\delta}_2\bigr)
          +\cos\!\bigl(\bm{k}\cdot\boldsymbol{\delta}_3\bigr)\right].
\end{align}
Here, $t^\prime$  and $t^{\prime\prime}$ are the amplitudes of the second- and third-NN hoppings, respectively.
Fig.~\ref{fig:kaome_lattice} shows the kagome lattice and a schematic illustration of the second- and third-NN hoppings.
These additional hopping terms respect all the symmetries of the conventional kagome lattice model.

\begin{figure}[tbp]
  \includegraphics[width=1.0\linewidth]{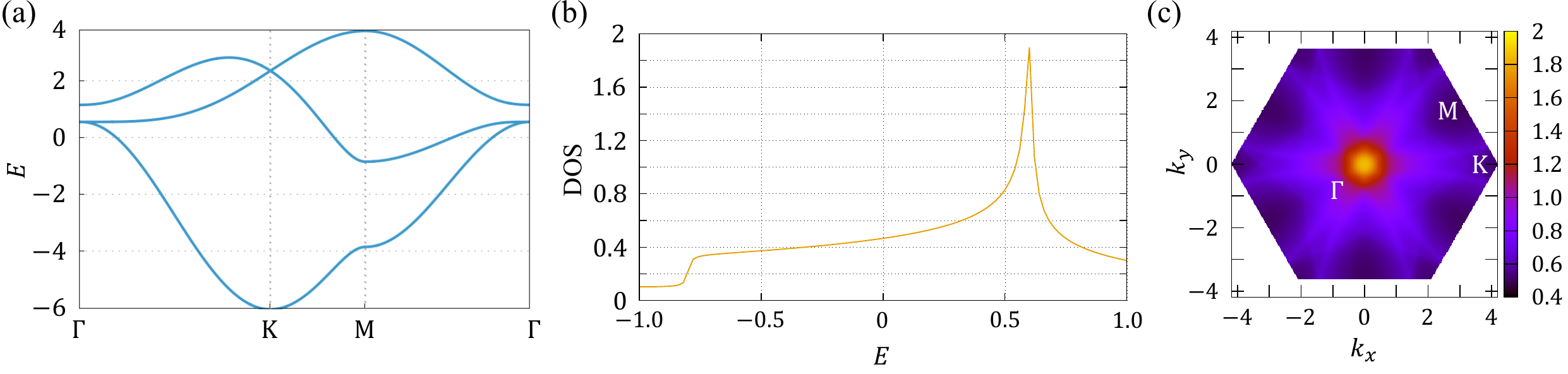}
  \centering
  \caption{(a) Energy dispersion, (b) DOS, and (c) noninteracting spin susceptibility of the extended kagome lattice for $(t, t^\prime, t^{\prime\prime}) = (1.00, -0.9, 0.20)$.
  }
  \label{fig:kaome}
\end{figure}

We demonstrate that the extended kagome lattice exhibits a singular saddle point by setting the parameters to $(t, t^\prime, t^{\prime\prime}) = (1, -0.9, 0.2)$.
Fig.~\ref{fig:kaome}(a) shows the corresponding energy dispersion. We find that the singular saddle point appears at the $\Gamma$ point. The effective mass of the dispersion along the $\Gamma$-$K$ line has the opposite sign to that along the $\Gamma$-$M$ line. In $C_4$-symmetric systems, a singular saddle point is characterized by
the intersection of two dispersions at an angle of $\pi/4$,
with opposite signs of the effective mass. By contrast, in the present case, two energy dispersions with opposite signs of effective mass intersect at $\pi/6$.

In our model, the singular saddle point occurs at the energy $E=0.6$. Fig.~\ref{fig:kaome}(b) shows the density of states calculated as $D(E)=\sum_{n}\int_{\rm BZ}\frac{d\bm k}{(2\pi)^2}\frac{\delta/\pi}{(E_n(\bm k)-E)^2+\delta^2}$ with $\delta=0.01$, where $E_n(\bm k)$ denotes the energy dispersion of band $n$. The resulting $D(E)$ exhibits a divergent behavior at $E=0.6$.
We also show in Fig.~\ref{fig:kaome}(c) the noninteracting spin susceptibility for the chemical potential $\mu = 0.6$ and the temperature $T = 0.01$, for which the singular saddle point lies at the Fermi energy.
The noninteracting spin susceptibility is calculated as,
\begin{align}
    \chi_0(\bm q) = \sum_{nm}\int\dfrac{d\bm k}{(2\pi)^2}
    \dfrac{f(E_m(\bm k))-f(E_n(\bm k+\bm q))}{E_n(\bm k+\bm q)-E_m(\bm k)}\vert\braket{u_n(\bm k+\bm q)\vert u_m(\bm k)}\vert^2,
\end{align}
where $\ket{u_n(\bm k)}$ is the Bloch wave function of band $n$ and $f(E)$ is the Fermi distribution function.
We find that the extended kagome lattice exhibits ferromagnetic correlations.
These results provide strong evidence for quantum geometric ferromagnetism
from a singular saddle point protected by $C_3$ symmetry.

\section{Construction of Quasi-local States for the $t_{2g}$-Orbital Model}
In the main text, we showed that the two-dimensional (2D) $t_{2g}$-orbital model hosts a singular flat band for an appropriate sets of parameters and constructed the quasi-local states (QLSs) associated with it. Such QLSs play an essential role in the proof of flat-band ferromagnetism~\cite{SMielke1999,STasaki2020}. In this section, we present the detailed construction of the QLSs for the flat-band 2D $t_{2g}$-orbital model.

\subsection{Flat-band 2D $t_{\rm 2g}$-orbital model}
We consider the 2D Cartesian coordinate system $\bm r$ with lattice unit vectors $\hat{\bm x} = (1, 0)$ and $\hat{\bm y} = (0,1)$.
We adopt the periodic boundary conditions and assume that the number of sites in the $x$ and $y$ directions is $L$. 
The total number of unit cells is given by $N = L^2$.
In this coordinate system, we introduce the $\beta$ operator
\begin{align}
    \hat{\beta}_{\sigma,\bm r} &= \hat{c}_{x,\sigma,\bm r} - \hat{c}_{x,\sigma,\bm r+\hat{\bm x}}
  + \hat{c}_{x,\sigma,\bm r+\hat{\bm y}} - \hat{c}_{x,\sigma,\bm r+\hat{\bm x} + \hat{\bm y}}
  \notag\\
  &+
  \hat{c}_{y,\sigma,\bm r}+ \hat{c}_{y,\sigma,\bm r+\hat{\bm x}}
  - \hat{c}_{y,\sigma,\bm r+\hat{\bm y}} - \hat{c}_{y,\sigma,\bm r+\hat{\bm x} + \hat{\bm y}},
\end{align}
where $\hat{c}_{l,\sigma, \bm r}$ is the annihilation operator with orbital $l=x,y$ and spin $\sigma=\uparrow,\downarrow$ at $\bm r$, and $\hat{N}$ is the number operator.
Here, $x$ and $y$ denote the $d_{ xz}$ and $d_{ yz}$ orbitals, respectively. 
The Hamiltonian of the flat-band 2D $t_{2g}$-orbital model on a square lattice is described by this $\beta$ operator as
\begin{eqnarray}
  \hat{H}_0 &=& \dfrac{t}{2}\sum_{\bm r,{\sigma}}\hat{\beta}^\dagger_{\sigma,\bm r}\hat{\beta}_{\sigma,\bm r} -2t\hat{N},~\label{eq:Hamiltonian_t2g}
\end{eqnarray}
with hopping amplitude $t$.
By taking $t = -1$, the energy dispersion of Fig.3 (d) in the main text is reproduced by the single-electron eigenvalue of $\hat{H}_0$.
Thus, as shown in the main text, $\hat{H}_0$ has a perfect flat band with band touching.

\subsection{Compact localized states and noncontractible loop states}

Hereafter, we focus on the up spin $\sigma=\uparrow$ and omit the spin degrees of freedom since the construction of QLSs for the down spin is the same as that for the up spin.
The $\beta$ operator anticommutes with the following $\alpha$ operators
\begin{align}
    \hat{\alpha}_{\bm r}=&
    \hat c_{x,\bm r}
    + \hat c_{x,\bm r+\hat{\bm x}}
    -\hat c_{x,\bm r+\hat{\bm y}}
    -\hat c_{x,\bm r+\hat{\bm x}+\hat{\bm y}}
    \nonumber\\
    &\quad
    -\hat c_{y,\bm r}
    + \hat c_{y,\bm r+\hat{\bm x}}
    -\hat c_{y,\bm r+\hat{\bm y}}
    + \hat c_{y,\bm r+\hat{\bm x}+\hat{\bm y}},
\end{align}
i.e. $\{\hat{\alpha}_{\bm r}^\dagger,\beta_{\bm r^\prime}\} = 0$.
This anticommutation relation indicates that the single-electron states $\ket{\alpha_{\bm r}} = \hat{\alpha}^\dagger_{\bm r}\ket{0}$ are the eigenstates of $\hat{H}_0$, i.e. $\hat{H}_0\ket{\alpha_{\bm r}} = -2t\ket{\alpha_{\bm r}}$, with the degenerate eigenvalue $-2t$, where $\ket{0}$ is the vacuum state.
These flat band eigenstates are so-called compact localized states (CLSs) since $\ket{\alpha_{\bm r}}$ are strictly localized on finite sites.

In the case of an isolated flat band, the basis of the flat-band eigenstates is spanned only by CLSs.
However, there exists a class of flat bands for which the basis of their eigenstates is not spanned only by CLSs due to the band touching~\cite{SBergman2008-nc,SRhim2019-km}. 
Instead, the basis of the flat-band eigenstates is given by a proper combination of CLSs and states that extend in one direction, namely, the noncontractible loop states (NLSs)~\cite{SBergman2008-nc,SRhim2019-km}.
Such a flat band is called a singular flat band, and our flat band with band touchings belongs to this class.

As shown in the main text, our flat band has two band touchings at $\Gamma$ and $M$ points.
To ensure that the $M$ point is included among the allowed $\bm k$ points, we assume that $L$ is an even number.
Therefore, the single-electron eigenstates of $\hat{H}_0$ are $N+2$ degenerate.
There are two types of NLSs that extend in two different directions for each band touching.
Two NLSs for the $\Gamma$ point are given by
\begin{align}
|\gamma_{x,r_y}\rangle &= \sum_{r_x}\hat c^\dagger_{x,(r_x,r_y)}|0\rangle, \qquad
|\gamma_{y,r_x}\rangle = \sum_{r_y}\hat c^\dagger_{y,(r_x,r_y)}|0\rangle, 
\end{align}
and two NLSs for the $M$ point are given by,
\begin{align}
    |\mu_{y,r_y}\rangle &= \sum_{r_x}(-1)^{r_x}\hat c^\dagger_{y,(r_x,r_y)}|0\rangle,\qquad
|\mu_{x,r_x}\rangle = \sum_{r_y}(-1)^{r_y}\hat c^\dagger_{x,(r_x,r_y)}|0\rangle .
\end{align}
A proper choice of $(N-2)$ CLSs and $4$ NLSs spans the basis of flat-band eigenstates.

\subsection{Quasi-local state}
While we can construct a basis of flat-band eigenstates by CLSs and NLSs, we here introduce an alternative basis of flat-band eigenstates, which is given by so-called quasi-local states (QLSs)~\cite{STasaki2020}.
The QLSs are strictly localized states with respect to a specific subset of sites (or orbitals). More precisely, when restricted to this subset, each QLS has a nonzero amplitude on only one element of the subset and vanishes on all the others.
Moreover, these distinguished elements are different for different QLSs, so that no two QLSs are associated with the same element of the subset.
Hence, the QLSs are in one-to-one correspondence with the elements of the subset.

The QLSs are useful for diagnosing the ground state of the lowest (highest) half-filled flat band of the Hubbard model.
If the QLSs satisfy the connectivity condition of Mielke's theorem for flat-band ferromagnetism, then the ground state of the lowest (highest) half-filled flat band exhibits ferromagnetism~\cite{SMielke1999,STasaki2020}.
In the following, we construct the QLSs of the flat-band 2D $t_{\rm 2g}$-orbital model.

\begin{figure}[tbp]
  \includegraphics[width=0.5\linewidth]{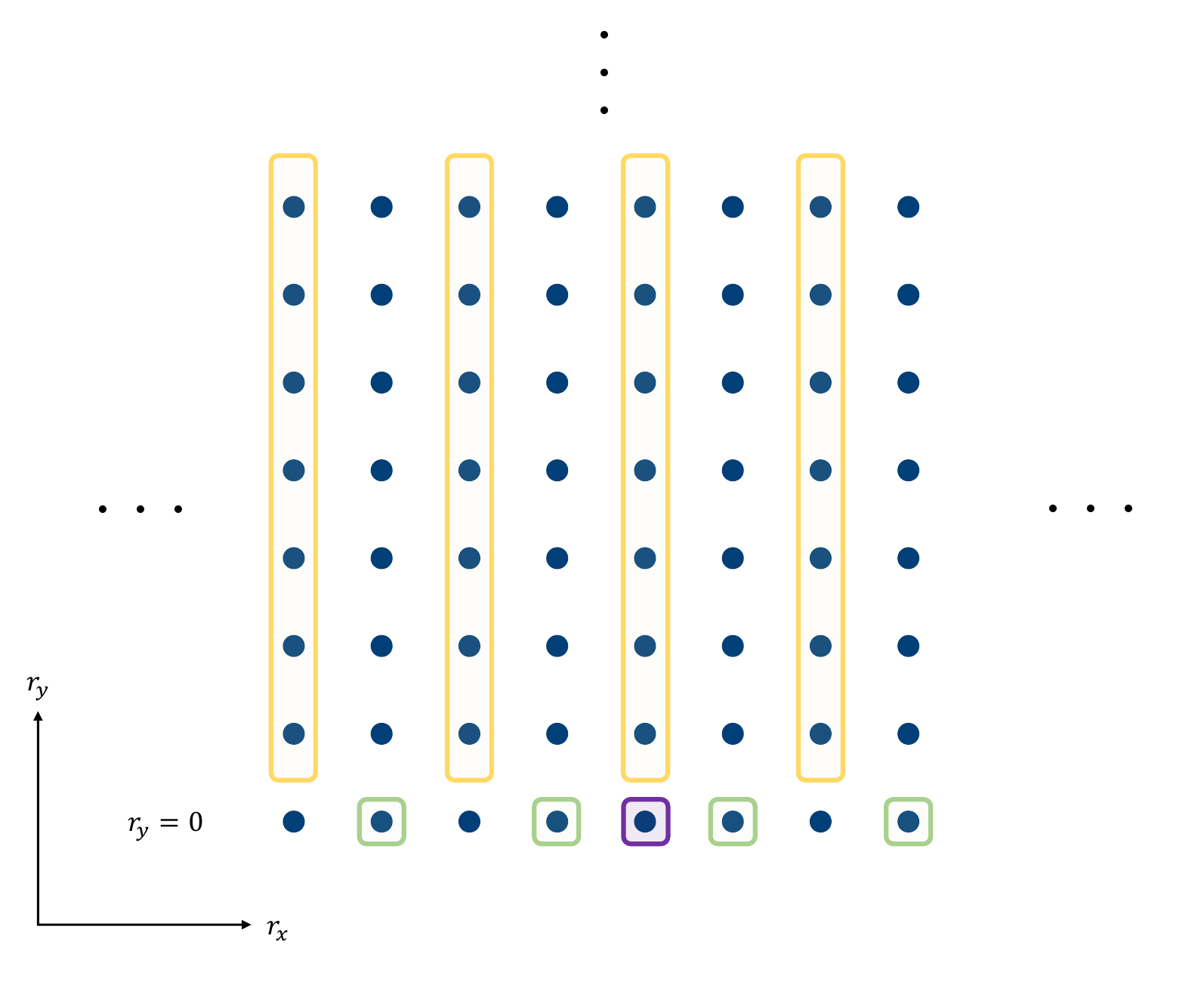}
  \centering
  \caption{Schematic illustration for the sites in $\Lambda_{r_1},\Lambda_{r_2}$ and $\Lambda_{r_3}$, which are highlighted by yellow, green, and purple colors, respectively.  The navy solid circles denote sites.  The lowest sites correspond to $r_y = 0$.
  }
  \label{fig:site}
\end{figure}

\subsubsection{Choice of orbital subset for QLSs}
Since our model has orbital degrees of freedom, i.e., $d_{ xz}$ and $d_{ yz}$ orbitals, each site has these two orbitals. Here, we choose the subset of orbitals $\Lambda_\rho$ where QLSs are strictly localized.
The total number of orbitals included in $\Lambda_\rho$ is $N+2$.
$N/2+1$ of them are given by $d_{xz}$ orbitals while the rest of them are given by $d_{yz}$ orbitals.
The subset consisting of the $d_{xz}$ ($d_{yz}$) orbitals is denoted by $\Lambda_x$ ($\Lambda_y$), which satisfies $\Lambda_\rho = \Lambda_x\cup \Lambda_y$.

The orbitals in $\Lambda_x$ and $\Lambda_y$ belong to the same set of sites.
These sites are classified into three types, which are given by
\begin{align}
    &\Lambda_{r_1} = \Big\{\bm r = (2n,m)\Big\vert
    n\in\{0,1,\cdots,L/2-1\},m\in\{1,2,\cdots,L-1\}\Big\},\\
    &\Lambda_{r_2} = \Big\{\bm r = (2n+1,0)\Big\vert
    n\in\{0,1,\cdots,L/2-1\}\Big\},\\
    &\Lambda_{r_3} = \Big\{\bm (r_x^*,0)\Big\},
\end{align}
where $r_x^*$ is fixed to a certain even number.
In Fig.~\ref{fig:site}, sites in $\Lambda_{r_1},\Lambda_{r_2}$ and $\Lambda_{r_3}$ are schematically shown by yellow, green, and purple-colored boxes.
By using these localized sites, $\Lambda_{l}$ for $l = {x,y}$ are defined by
\begin{align}
        \Lambda_l &= \Lambda_{l_1}\cup\Lambda_{l_2}\cup\Lambda_{l_3},\\
    \Lambda_{l_i}  
    &= \Big\{\tau_l = (d_{l z},\bm r)\Big\vert \bm r\in\Lambda_{r_i}\Big\},
\end{align}
with $i = 1,2,3$.
Thus, each QLS is labeled by $\tau_{l}\in\Lambda_{l}$ and we denote it by $\ket{\rho_{\tau_{l}}}$.

\subsubsection{QLSs for $\tau_{l}\in\Lambda_{l_1}$}
\begin{figure}[tbp]
  \includegraphics[width=0.5\linewidth]{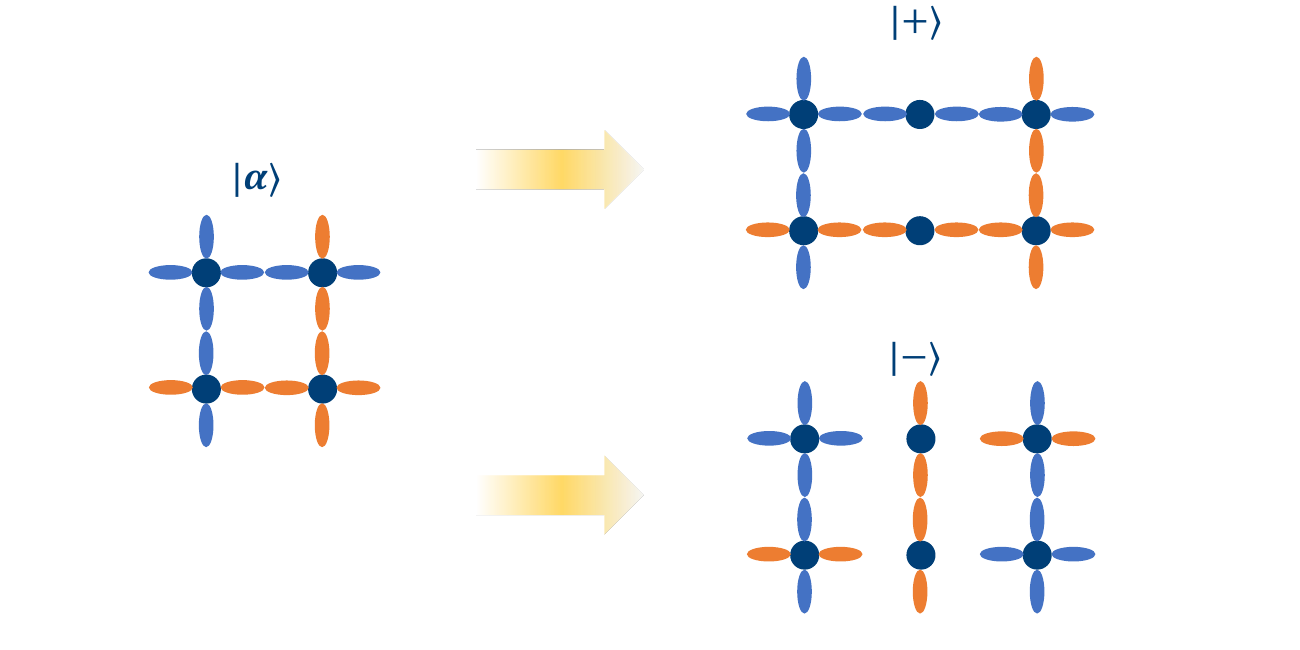}
  \centering
  \caption{Schematic illustration for $\ket{\alpha_{\bm r}}$ and $\ket{\pm_{\bm r}}$. The two ellipses on either side of the navy-blue circle, spreading horizontally (vertically), represent the $d_{xz}$ ($d_{yz}$) orbitals. The color of orbitals indicates the phase of the orbital in each state. The red and blue indicate positive and negative signs, respectively.
  }
  \label{fig:building}
\end{figure}

We construct the QLSs for $\tau_l\in\Lambda_{l_1}$.
As building blocks of QLSs, we introduce two different states constructed by $\ket{\alpha_{\bm r}}$ and $\ket{\alpha_{{\bm r}-\hat{\bm x}}}$ as
\begin{align}
    \ket{\pm_{\bm r}} = \pm\ket{\alpha_{\bm r}}+\ket{\alpha_{\bm r-\hat{\bm x}}}.
\end{align}
The states $\ket{\alpha_{\bm r}}$ and $\ket{\pm_{\bm r}}$ are schematically illustrated in Fig.~\ref{fig:building}.
Only $d_{xz}$ and $d_{yz}$ orbitals have nonzero amplitudes at the middle sites of $\ket{+_{\bm r}}$ and $\ket{-_{\bm r}}$, respectively.

\begin{figure}[tbp]
  \includegraphics[width=1.0\linewidth]{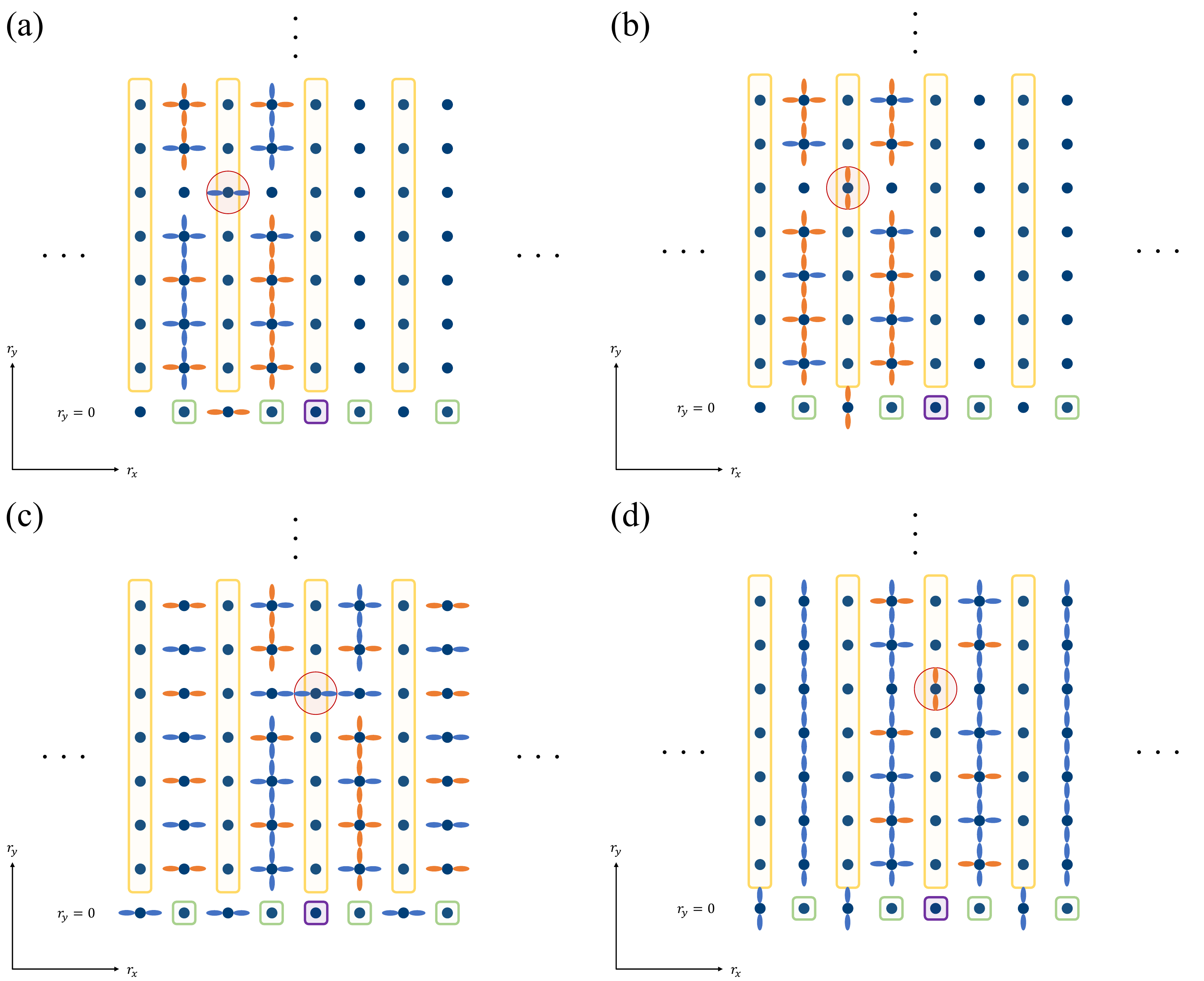}
  \centering
  \caption{Schematic illustration of the QLSs for (a,c) $\tau_{x}\in\Lambda_{x_1}$ and (b,d) $\tau_{y}\in\Lambda_{y_1}$.
  In each figure, the red circle highlights the strictly localized orbital.
  }
  \label{fig:QLSs_first}
\end{figure}

The basic strategy for constructing QLSs is as follows. First, we construct strictly localized states with respect to $\Lambda_{l_1}$ by using $\ket{\pm_{\bm r}}$. Since these states have nonzero amplitude at the sites in $\Lambda_{l_2}$ and $\Lambda_{l_3}$, we then cancel such residual amplitudes by adding appropriate NLSs.
For $r_x\neq r_x^*$, the resulting QLSs are given by,
\begin{align}
    \ket{\rho_{\tau_x}} &= \ket{\phi_{\bm r}}, \qquad
    \ket{\phi_{\bm r}} \equiv
    \sum_{r_y^\prime = 0}^{r_y{-1}}\ket{+_{(r_x,r_y^\prime)}}
    + \ket{\gamma_{y,r_x-1}} - \ket{\gamma_{y,r_x+1}}
    - \ket{\mu_{x,r_x-1}} - \ket{\mu_{x,r_x+1}}, \\
    \ket{\rho_{\tau_y}} &= \ket{\psi_{\bm r}}, \qquad
    \ket{\psi_{\bm r}} \equiv
    \sum_{r_y^\prime = 0}^{r_y{-1}}(-1)^{r_y^\prime}\ket{-_{(r_x,r_y^\prime)}}
    + \ket{\gamma_{y,r_x-1}} + \ket{\gamma_{y,r_x+1}}
    - \ket{\mu_{x,r_x-1}} + \ket{\mu_{x,r_x+1}} .
\end{align}
For the latter construction of QLSs, we introduce $\ket{\phi_{\bm r}}$ and $\ket{\psi_{\bm r}}$, which are schematically illustrated in Figs.~\ref{fig:QLSs_first} (a) and (b), respectively.

We find that these states have finite amplitude at $(r_x,0)$.
Therefore, for $r_x = r_x^*$, $\ket{\phi_{\bm r}}$ and $\ket{\psi_{\bm r}}$ themselves are not QLSs since $(r_x^*,0)$ is included in $\Lambda_{l_3}$.
In this case, we further add NLSs to $\ket{\phi_{\bm r}}$ and $\ket{\psi_{\bm r}}$ and the resulting QLSs are given by,
\begin{align}
    &\ket{\rho_{\tau_x}} = 
    \ket{\phi_{\bm r}} + 2\sum_{n = 1}^{L/2}\ket{\mu_{x,2n-1}}
    -2\ket{\gamma_{x,0}}, \\
    &\ket{\rho_{\tau_y}} = 
    \ket{\psi_{\bm r}} {-} 2\sum_{n = {1}}^{L/2}\ket{\gamma_{{y},2n-1}}
    -2\ket{\mu_{y,0}},
\end{align}
which are schematically illustrated in Figs.~\ref{fig:QLSs_first} (c) and (d), respectively.

\subsubsection{QLSs for $\tau_{l}\in\Lambda_{l_2}$}
\begin{figure}[tbp]
  \includegraphics[width=1.0\linewidth]{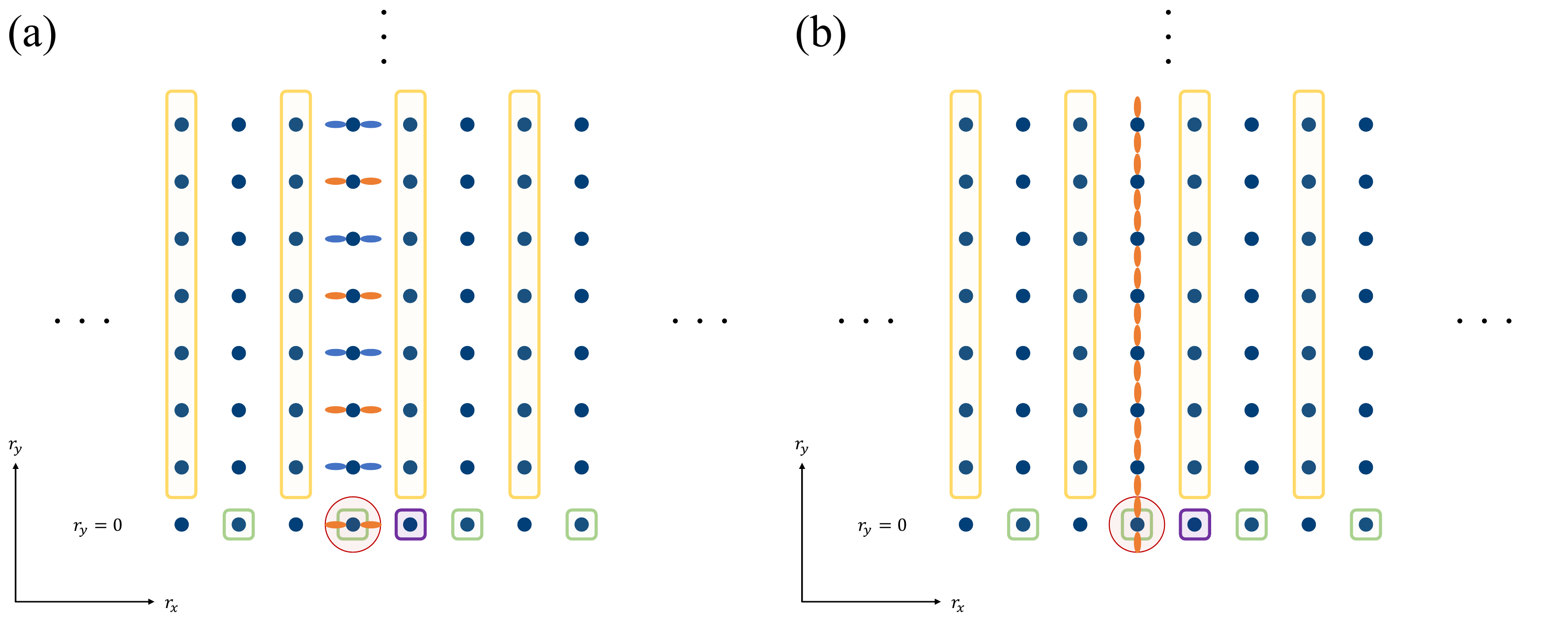}
  \centering
  \caption{Schematic illustration of the QLSs for (a) $\tau_{x}\in\Lambda_{x_2}$ and (b) $\tau_{y}\in\Lambda_{y_2}$.
  }
  \label{fig:QLSs_second}
\end{figure}

QLSs for $\tau_{l}\in\Lambda_{l_2}$ are simply given by NLSs themselves as
\begin{align}
    &\ket{\rho_{\tau_x}} = 
    \ket{\mu_{x,r_x}}, 
    \\
    &\ket{\rho_{\tau_y}} = 
    \ket{\gamma_{y,r_x}}, 
\end{align}
which are shown in Figs.~\ref{fig:QLSs_second} (a) and (b), respectively.

\subsubsection{QLSs for $\tau_{l}\in\Lambda_{l_3}$}
\begin{figure}[tbp]
  \includegraphics[width=1.0\linewidth]{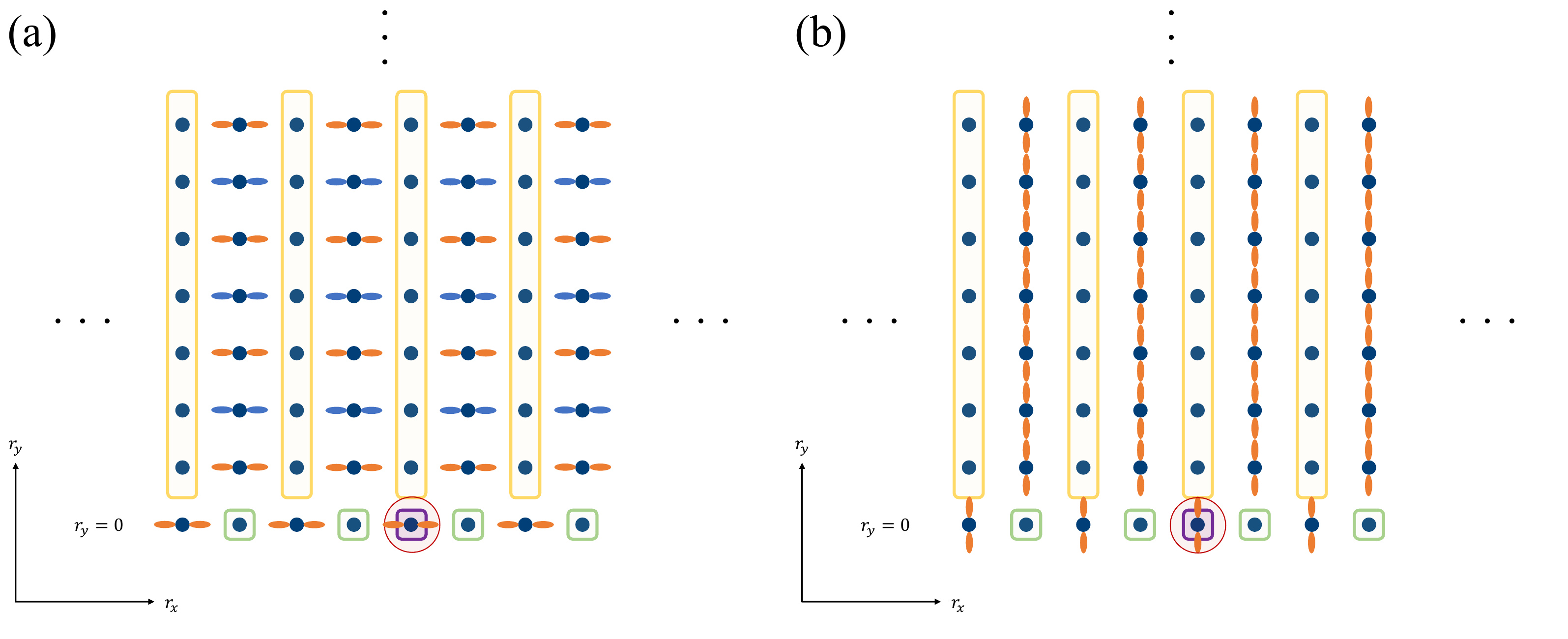}
  \centering
  \caption{Schematic illustration of the QLSs for (a) $\tau_{x}\in\Lambda_{x_3}$ and (b) $\tau_{y}\in\Lambda_{y_3}$.
  }
  \label{fig:QLSs_third}
\end{figure}

For $\tau_l\in\Lambda_{l_3}$, we start from $\ket{\gamma_{x,0}}$ and $\ket{\mu_{y,0}}$, both of which extend in the $x$ direction. Since these states have amplitudes on sites in both $\Lambda_{l_2}$ and $\Lambda_{l_3}$, we cancel the residual amplitudes on $\Lambda_{l_2}$ by adding appropriate NLSs extending in the $y$ direction; the resulting QLSs are given by,
\begin{align}
    &\ket{\rho_{\tau_x}} = 
    \ket{\gamma_{x,0}} - \sum_{n = 1}^{L/2}\ket{\mu_{x,2n-1}},\\
    &\ket{\rho_{\tau_y}} =
    \ket{\mu_{y,0}} + \sum_{n = {1}}^{L/2}\ket{\gamma_{y,2n-1}},
\end{align}
which are schematically illustrated in Figs.~\ref{fig:QLSs_third} (a) and (b), respectively.

\begin{thebibliography}{78}%
\makeatletter
\providecommand \@ifxundefined [1]{%
 \@ifx{#1\undefined}
}%
\providecommand \@ifnum [1]{%
 \ifnum #1\expandafter \@firstoftwo
 \else \expandafter \@secondoftwo
 \fi
}%
\providecommand \@ifx [1]{%
 \ifx #1\expandafter \@firstoftwo
 \else \expandafter \@secondoftwo
 \fi
}%
\providecommand \natexlab [1]{#1}%
\providecommand \enquote  [1]{``#1''}%
\providecommand \bibnamefont  [1]{#1}%
\providecommand \bibfnamefont [1]{#1}%
\providecommand \citenamefont [1]{#1}%
\providecommand \href@noop [0]{\@secondoftwo}%
\providecommand \href [0]{\begingroup \@sanitize@url \@href}%
\providecommand \@href[1]{\@@startlink{#1}\@@href}%
\providecommand \@@href[1]{\endgroup#1\@@endlink}%
\providecommand \@sanitize@url [0]{\catcode `\\12\catcode `\$12\catcode
  `\&12\catcode `\#12\catcode `\^12\catcode `\_12\catcode `\%12\relax}%
\providecommand \@@startlink[1]{}%
\providecommand \@@endlink[0]{}%
\providecommand \url  [0]{\begingroup\@sanitize@url \@url }%
\providecommand \@url [1]{\endgroup\@href {#1}{\urlprefix }}%
\providecommand \urlprefix  [0]{URL }%
\providecommand \Eprint [0]{\href }%
\providecommand \doibase [0]{https://doi.org/}%
\providecommand \selectlanguage [0]{\@gobble}%
\providecommand \bibinfo  [0]{\@secondoftwo}%
\providecommand \bibfield  [0]{\@secondoftwo}%
\providecommand \translation [1]{[#1]}%
\providecommand \BibitemOpen [0]{}%
\providecommand \bibitemStop [0]{}%
\providecommand \bibitemNoStop [0]{.\EOS\space}%
\providecommand \EOS [0]{\spacefactor3000\relax}%
\providecommand \BibitemShut  [1]{\csname bibitem#1\endcsname}%
\let\auto@bib@innerbib\@empty
\bibitem [{\citenamefont {Moriya}()}]{Moriya1985}%
  \BibitemOpen
  \bibfield  {author} {\bibinfo {author} {\bibfnamefont {T.}~\bibnamefont
  {Moriya}},\ }\href
  {https://doi.org/https://doi.org/10.1007/978-3-642-82499-9} {\bibinfo {title}
  {Spin fluctuations in itinerant electron magnetism}},\ \bibinfo {note}
  {(Springer, 1985)}\BibitemShut {NoStop}%
\bibitem [{\citenamefont {Terakura}\ \emph {et~al.}(1982)\citenamefont
  {Terakura}, \citenamefont {Hamada}, \citenamefont {Oguchi},\ and\
  \citenamefont {Asada}}]{Terakura1982}%
  \BibitemOpen
  \bibfield  {author} {\bibinfo {author} {\bibfnamefont {K.}~\bibnamefont
  {Terakura}}, \bibinfo {author} {\bibfnamefont {N.}~\bibnamefont {Hamada}},
  \bibinfo {author} {\bibfnamefont {T.}~\bibnamefont {Oguchi}},\ and\ \bibinfo
  {author} {\bibfnamefont {T.}~\bibnamefont {Asada}},\ }\href
  {https://iopscience.iop.org/article/10.1088/0305-4608/12/8/012} {\bibfield
  {journal} {\bibinfo  {journal} {J. Phys. \TK{F: Met. Phys.}}\ }\textbf
  {\bibinfo {volume} {12}},\ \bibinfo {pages} {1661} (\bibinfo {year}
  {1982})}\BibitemShut {NoStop}%
\bibitem [{\citenamefont {Saxena}\ \emph {et~al.}(2000)\citenamefont {Saxena},
  \citenamefont {Agarwal}, \citenamefont {Ahilan}, \citenamefont {Grosche},
  \citenamefont {Haselwimmer}, \citenamefont {Steiner}, \citenamefont {Pugh},
  \citenamefont {Walker}, \citenamefont {Julian}, \citenamefont {Monthoux},
  \citenamefont {Lonzarich}, \citenamefont {Huxley}, \citenamefont {Sheikin},
  \citenamefont {Braithwaite},\ and\ \citenamefont {Flouquet}}]{Saxena2000}%
  \BibitemOpen
  \bibfield  {author} {\bibinfo {author} {\bibfnamefont {S.~S.}\ \bibnamefont
  {Saxena}}, \bibinfo {author} {\bibfnamefont {P.}~\bibnamefont {Agarwal}},
  \bibinfo {author} {\bibfnamefont {K.}~\bibnamefont {Ahilan}}, \bibinfo
  {author} {\bibfnamefont {F.~M.}\ \bibnamefont {Grosche}}, \bibinfo {author}
  {\bibfnamefont {R.~K.~{\TK{W}}.}\ \bibnamefont {Haselwimmer}}, \bibinfo
  {author} {\bibfnamefont {M.~J.}\ \bibnamefont {Steiner}}, \bibinfo {author}
  {\bibfnamefont {E.}~\bibnamefont {Pugh}}, \bibinfo {author} {\bibfnamefont
  {I.~R.}\ \bibnamefont {Walker}}, \bibinfo {author} {\bibfnamefont {S.~R.}\
  \bibnamefont {Julian}}, \bibinfo {author} {\bibfnamefont {P.}~\bibnamefont
  {Monthoux}}, \bibinfo {author} {\bibfnamefont {G.~G.}\ \bibnamefont
  {Lonzarich}}, \bibinfo {author} {\bibfnamefont {A.}~\bibnamefont {Huxley}},
  \bibinfo {author} {\bibfnamefont {I.}~\bibnamefont {Sheikin}}, \bibinfo
  {author} {\bibfnamefont {D.}~\bibnamefont {Braithwaite}},\ and\ \bibinfo
  {author} {\bibfnamefont {J.}~\bibnamefont {Flouquet}},\ }\href
  {https://www.nature.com/articles/35020500} {\bibfield  {journal} {\bibinfo
  {journal} {Nature}\ }\textbf {\bibinfo {volume} {406}},\ \bibinfo {pages}
  {587} (\bibinfo {year} {2000})}\BibitemShut {NoStop}%
\bibitem [{\citenamefont {Aoki}\ \emph {et~al.}(2001)\citenamefont {Aoki},
  \citenamefont {Huxley}, \citenamefont {Ressouche}, \citenamefont
  {Braithwaite}, \citenamefont {Flouquet}, \citenamefont {Brison},
  \citenamefont {Lhotel},\ and\ \citenamefont {Paulsen}}]{Aoki2001}%
  \BibitemOpen
  \bibfield  {author} {\bibinfo {author} {\bibfnamefont {D.}~\bibnamefont
  {Aoki}}, \bibinfo {author} {\bibfnamefont {A.}~\bibnamefont {Huxley}},
  \bibinfo {author} {\bibfnamefont {E.}~\bibnamefont {Ressouche}}, \bibinfo
  {author} {\bibfnamefont {D.}~\bibnamefont {Braithwaite}}, \bibinfo {author}
  {\bibfnamefont {J.}~\bibnamefont {Flouquet}}, \bibinfo {author}
  {\bibfnamefont {J.~P.}\ \bibnamefont {Brison}}, \bibinfo {author}
  {\bibfnamefont {E.}~\bibnamefont {Lhotel}},\ and\ \bibinfo {author}
  {\bibfnamefont {C.}~\bibnamefont {Paulsen}},\ }\href
  {https://www.nature.com/articles/35098048} {\bibfield  {journal} {\bibinfo
  {journal} {Nature}\ }\textbf {\bibinfo {volume} {413}},\ \bibinfo {pages}
  {613} (\bibinfo {year} {2001})}\BibitemShut {NoStop}%
\bibitem [{\citenamefont {Huy}\ \emph {et~al.}(2007)\citenamefont {Huy},
  \citenamefont {Gasparini}, \citenamefont {de~Nijs}, \citenamefont {Huang},
  \citenamefont {Klaasse}, \citenamefont {Gortenmulder}, \citenamefont
  {de~Visser}, \citenamefont {Hamann}, \citenamefont {Görlach},\ and\
  \citenamefont {Löhneysen}}]{Huy2007}%
  \BibitemOpen
  \bibfield  {author} {\bibinfo {author} {\bibfnamefont {N.~T.}\ \bibnamefont
  {Huy}}, \bibinfo {author} {\bibfnamefont {A.}~\bibnamefont {Gasparini}},
  \bibinfo {author} {\bibfnamefont {D.~E.}\ \bibnamefont {de~Nijs}}, \bibinfo
  {author} {\bibfnamefont {Y.}~\bibnamefont {Huang}}, \bibinfo {author}
  {\bibfnamefont {J.~C.~P.}\ \bibnamefont {Klaasse}}, \bibinfo {author}
  {\bibfnamefont {T.}~\bibnamefont {Gortenmulder}}, \bibinfo {author}
  {\bibfnamefont {A.}~\bibnamefont {de~Visser}}, \bibinfo {author}
  {\bibfnamefont {A.}~\bibnamefont {Hamann}}, \bibinfo {author} {\bibfnamefont
  {T.}~\bibnamefont {Görlach}},\ and\ \bibinfo {author} {\bibfnamefont
  {H.~{\TK{v}}.}\ \bibnamefont {Löhneysen}},\ }\href
  {https://journals.aps.org/prl/abstract/10.1103/PhysRevLett.99.067006}
  {\bibfield  {journal} {\bibinfo  {journal} {Phys. Rev. Lett.}\ }\textbf
  {\bibinfo {volume} {99}},\ \bibinfo {pages} {067006} (\bibinfo {year}
  {2007})}\BibitemShut {NoStop}%
\bibitem [{\citenamefont {Aoki}\ \emph {et~al.}(2019)\citenamefont {Aoki},
  \citenamefont {Ishida},\ and\ \citenamefont {Flouquet}}]{Aoki2019}%
  \BibitemOpen
  \bibfield  {author} {\bibinfo {author} {\bibfnamefont {D.}~\bibnamefont
  {Aoki}}, \bibinfo {author} {\bibfnamefont {K.}~\bibnamefont {Ishida}},\ and\
  \bibinfo {author} {\bibfnamefont {J.}~\bibnamefont {Flouquet}},\ }\href
  {https://journals.jps.jp/doi/10.7566/JPSJ.88.022001} {\bibfield  {journal}
  {\bibinfo  {journal} {J. Phys. Soc. Jpn.}\ }\textbf {\bibinfo {volume}
  {88}},\ \bibinfo {pages} {022001} (\bibinfo {year} {2019})}\BibitemShut
  {NoStop}%
\bibitem [{\citenamefont {Burch}\ \emph {et~al.}(2018)\citenamefont {Burch},
  \citenamefont {Mandrus},\ and\ \citenamefont {Park}}]{Burch2018}%
  \BibitemOpen
  \bibfield  {author} {\bibinfo {author} {\bibfnamefont {K.~S.}\ \bibnamefont
  {Burch}}, \bibinfo {author} {\bibfnamefont {D.}~\bibnamefont {Mandrus}},\
  and\ \bibinfo {author} {\bibfnamefont {J.-G.}\ \bibnamefont {Park}},\ }\href
  {https://www.nature.com/articles/s41586-018-0631-z} {\bibfield  {journal}
  {\bibinfo  {journal} {Nature}\ }\textbf {\bibinfo {volume} {563}},\ \bibinfo
  {pages} {47} (\bibinfo {year} {2018})}\BibitemShut {NoStop}%
\bibitem [{\citenamefont {Gong}\ and\ \citenamefont {Zhang}(2019)}]{Gong2019}%
  \BibitemOpen
  \bibfield  {author} {\bibinfo {author} {\bibfnamefont {C.}~\bibnamefont
  {Gong}}\ and\ \bibinfo {author} {\bibfnamefont {X.}~\bibnamefont {Zhang}},\
  }\href {https://doi.org/10.1126/science.aav4450} {\bibfield  {journal}
  {\bibinfo  {journal} {Science}\ }\textbf {\bibinfo {volume} {363}},\ \bibinfo
  {pages} {eaav4450} (\bibinfo {year} {2019})}\BibitemShut {NoStop}%
\bibitem [{\citenamefont {Wang}\ \emph {et~al.}(2022)\citenamefont {Wang},
  \citenamefont {Bedoya-Pinto}, \citenamefont {Blei}, \citenamefont {Dismukes},
  \citenamefont {Hamo}, \citenamefont {Jenkins}, \citenamefont {Koperski},
  \citenamefont {Liu}, \citenamefont {Sun}, \citenamefont {Telford},
  \citenamefont {Kim}, \citenamefont {Augustin}, \citenamefont {Vool},
  \citenamefont {Yin}, \citenamefont {Li}, \citenamefont {Falin}, \citenamefont
  {Dean}, \citenamefont {Casanova}, \citenamefont {Evans}, \citenamefont
  {Chshiev}, \citenamefont {Mishchenko}, \citenamefont {Petrovic},
  \citenamefont {He}, \citenamefont {Zhao}, \citenamefont {Tsen}, \citenamefont
  {Gerardot}, \citenamefont {Brotons-Gisbert}, \citenamefont {Guguchia},
  \citenamefont {Roy}, \citenamefont {Tongay}, \citenamefont {Wang},
  \citenamefont {Hasan}, \citenamefont {Wrachtrup}, \citenamefont {Yacoby},
  \citenamefont {Fert}, \citenamefont {Parkin}, \citenamefont {Novoselov},
  \citenamefont {Dai}, \citenamefont {Balicas},\ and\ \citenamefont
  {Santos}}]{Wang2022}%
  \BibitemOpen
  \bibfield  {author} {\bibinfo {author} {\bibfnamefont {Q.~H.}\ \bibnamefont
  {Wang}}, \bibinfo {author} {\bibfnamefont {A.}~\bibnamefont {Bedoya-Pinto}},
  \bibinfo {author} {\bibfnamefont {M.}~\bibnamefont {Blei}}, \bibinfo {author}
  {\bibfnamefont {A.~H.}\ \bibnamefont {Dismukes}}, \bibinfo {author}
  {\bibfnamefont {A.}~\bibnamefont {Hamo}}, \bibinfo {author} {\bibfnamefont
  {S.}~\bibnamefont {Jenkins}}, \bibinfo {author} {\bibfnamefont
  {M.}~\bibnamefont {Koperski}}, \bibinfo {author} {\bibfnamefont
  {Y.}~\bibnamefont {Liu}}, \bibinfo {author} {\bibfnamefont {Q.-C.}\
  \bibnamefont {Sun}}, \bibinfo {author} {\bibfnamefont {E.~J.}\ \bibnamefont
  {Telford}}, \bibinfo {author} {\bibfnamefont {H.~H.}\ \bibnamefont {Kim}},
  \bibinfo {author} {\bibfnamefont {M.}~\bibnamefont {Augustin}}, \bibinfo
  {author} {\bibfnamefont {U.}~\bibnamefont {Vool}}, \bibinfo {author}
  {\bibfnamefont {J.-X.}\ \bibnamefont {Yin}}, \bibinfo {author} {\bibfnamefont
  {L.~H.}\ \bibnamefont {Li}}, \bibinfo {author} {\bibfnamefont
  {A.}~\bibnamefont {Falin}}, \bibinfo {author} {\bibfnamefont {C.~R.}\
  \bibnamefont {Dean}}, \bibinfo {author} {\bibfnamefont {F.}~\bibnamefont
  {Casanova}}, \bibinfo {author} {\bibfnamefont {R.~F.~L.}\ \bibnamefont
  {Evans}}, \bibinfo {author} {\bibfnamefont {M.}~\bibnamefont {Chshiev}},
  \bibinfo {author} {\bibfnamefont {A.}~\bibnamefont {Mishchenko}}, \bibinfo
  {author} {\bibfnamefont {C.}~\bibnamefont {Petrovic}}, \bibinfo {author}
  {\bibfnamefont {R.}~\bibnamefont {He}}, \bibinfo {author} {\bibfnamefont
  {L.}~\bibnamefont {Zhao}}, \bibinfo {author} {\bibfnamefont {A.~W.}\
  \bibnamefont {Tsen}}, \bibinfo {author} {\bibfnamefont {B.~D.}\ \bibnamefont
  {Gerardot}}, \bibinfo {author} {\bibfnamefont {M.}~\bibnamefont
  {Brotons-Gisbert}}, \bibinfo {author} {\bibfnamefont {Z.}~\bibnamefont
  {Guguchia}}, \bibinfo {author} {\bibfnamefont {X.}~\bibnamefont {Roy}},
  \bibinfo {author} {\bibfnamefont {S.}~\bibnamefont {Tongay}}, \bibinfo
  {author} {\bibfnamefont {Z.}~\bibnamefont {Wang}}, \bibinfo {author}
  {\bibfnamefont {M.~Z.}\ \bibnamefont {Hasan}}, \bibinfo {author}
  {\bibfnamefont {J.}~\bibnamefont {Wrachtrup}}, \bibinfo {author}
  {\bibfnamefont {A.}~\bibnamefont {Yacoby}}, \bibinfo {author} {\bibfnamefont
  {A.}~\bibnamefont {Fert}}, \bibinfo {author} {\bibfnamefont {S.}~\bibnamefont
  {Parkin}}, \bibinfo {author} {\bibfnamefont {K.~S.}\ \bibnamefont
  {Novoselov}}, \bibinfo {author} {\bibfnamefont {P.}~\bibnamefont {Dai}},
  \bibinfo {author} {\bibfnamefont {L.}~\bibnamefont {Balicas}},\ and\ \bibinfo
  {author} {\bibfnamefont {E.~J.~G.}\ \bibnamefont {Santos}},\ }\href
  {https://pubs.acs.org/doi/10.1021/acsnano.1c09150} {\bibfield  {journal}
  {\bibinfo  {journal} {ACS Nano}\ }\textbf {\bibinfo {volume} {16}},\ \bibinfo
  {pages} {6960} (\bibinfo {year} {2022})}\BibitemShut {NoStop}%
\bibitem [{\citenamefont {Gong}\ \emph {et~al.}(2017)\citenamefont {Gong},
  \citenamefont {Li}, \citenamefont {Li}, \citenamefont {Ji}, \citenamefont
  {Stern}, \citenamefont {Xia}, \citenamefont {Cao}, \citenamefont {Bao},
  \citenamefont {Wang}, \citenamefont {Wang}, \citenamefont {Qiu},
  \citenamefont {Cava}, \citenamefont {Louie}, \citenamefont {Xia},\ and\
  \citenamefont {Zhang}}]{Gong2017}%
  \BibitemOpen
  \bibfield  {author} {\bibinfo {author} {\bibfnamefont {C.}~\bibnamefont
  {Gong}}, \bibinfo {author} {\bibfnamefont {L.}~\bibnamefont {Li}}, \bibinfo
  {author} {\bibfnamefont {Z.}~\bibnamefont {Li}}, \bibinfo {author}
  {\bibfnamefont {H.}~\bibnamefont {Ji}}, \bibinfo {author} {\bibfnamefont
  {A.}~\bibnamefont {Stern}}, \bibinfo {author} {\bibfnamefont
  {Y.}~\bibnamefont {Xia}}, \bibinfo {author} {\bibfnamefont {T.}~\bibnamefont
  {Cao}}, \bibinfo {author} {\bibfnamefont {W.}~\bibnamefont {Bao}}, \bibinfo
  {author} {\bibfnamefont {C.}~\bibnamefont {Wang}}, \bibinfo {author}
  {\bibfnamefont {Y.}~\bibnamefont {Wang}}, \bibinfo {author} {\bibfnamefont
  {Z.~Q.}\ \bibnamefont {Qiu}}, \bibinfo {author} {\bibfnamefont {R.~J.}\
  \bibnamefont {Cava}}, \bibinfo {author} {\bibfnamefont {S.~G.}\ \bibnamefont
  {Louie}}, \bibinfo {author} {\bibfnamefont {J.}~\bibnamefont {Xia}},\ and\
  \bibinfo {author} {\bibfnamefont {X.}~\bibnamefont {Zhang}},\ }\href
  {https://doi.org/10.1038/nature22060} {\bibfield  {journal} {\bibinfo
  {journal} {Nature}\ }\textbf {\bibinfo {volume} {546}},\ \bibinfo {pages}
  {265} (\bibinfo {year} {2017})}\BibitemShut {NoStop}%
\bibitem [{\citenamefont {Huang}\ \emph {et~al.}(2017)\citenamefont {Huang},
  \citenamefont {Clark}, \citenamefont {Navarro-Moratalla}, \citenamefont
  {Klein}, \citenamefont {Cheng}, \citenamefont {Seyler}, \citenamefont
  {Zhong}, \citenamefont {Schmidgall}, \citenamefont {McGuire}, \citenamefont
  {Cobden}, \citenamefont {Yao}, \citenamefont {Xiao}, \citenamefont
  {Jarillo-Herrero},\ and\ \citenamefont {Xu}}]{Huang2017}%
  \BibitemOpen
  \bibfield  {author} {\bibinfo {author} {\bibfnamefont {B.}~\bibnamefont
  {Huang}}, \bibinfo {author} {\bibfnamefont {G.}~\bibnamefont {Clark}},
  \bibinfo {author} {\bibfnamefont {E.}~\bibnamefont {Navarro-Moratalla}},
  \bibinfo {author} {\bibfnamefont {D.~R.}\ \bibnamefont {Klein}}, \bibinfo
  {author} {\bibfnamefont {R.}~\bibnamefont {Cheng}}, \bibinfo {author}
  {\bibfnamefont {K.~L.}\ \bibnamefont {Seyler}}, \bibinfo {author}
  {\bibfnamefont {D.}~\bibnamefont {Zhong}}, \bibinfo {author} {\bibfnamefont
  {E.}~\bibnamefont {Schmidgall}}, \bibinfo {author} {\bibfnamefont {M.~A.}\
  \bibnamefont {McGuire}}, \bibinfo {author} {\bibfnamefont {D.~H.}\
  \bibnamefont {Cobden}}, \bibinfo {author} {\bibfnamefont {W.}~\bibnamefont
  {Yao}}, \bibinfo {author} {\bibfnamefont {D.}~\bibnamefont {Xiao}}, \bibinfo
  {author} {\bibfnamefont {P.}~\bibnamefont {Jarillo-Herrero}},\ and\ \bibinfo
  {author} {\bibfnamefont {X.}~\bibnamefont {Xu}},\ }\href
  {https://doi.org/10.1038/nature22391} {\bibfield  {journal} {\bibinfo
  {journal} {Nature}\ }\textbf {\bibinfo {volume} {546}},\ \bibinfo {pages}
  {270} (\bibinfo {year} {2017})}\BibitemShut {NoStop}%
\bibitem [{\citenamefont {Deng}\ \emph {et~al.}(2018)\citenamefont {Deng},
  \citenamefont {Yu}, \citenamefont {Song}, \citenamefont {Zhang},
  \citenamefont {Wang}, \citenamefont {Sun}, \citenamefont {Yi}, \citenamefont
  {Wu}, \citenamefont {Wu}, \citenamefont {Zhu}, \citenamefont {Wang},
  \citenamefont {Chen},\ and\ \citenamefont {Zhang}}]{Deng2018}%
  \BibitemOpen
  \bibfield  {author} {\bibinfo {author} {\bibfnamefont {Y.}~\bibnamefont
  {Deng}}, \bibinfo {author} {\bibfnamefont {Y.}~\bibnamefont {Yu}}, \bibinfo
  {author} {\bibfnamefont {Y.}~\bibnamefont {Song}}, \bibinfo {author}
  {\bibfnamefont {J.}~\bibnamefont {Zhang}}, \bibinfo {author} {\bibfnamefont
  {N.~Z.}\ \bibnamefont {Wang}}, \bibinfo {author} {\bibfnamefont
  {Z.}~\bibnamefont {Sun}}, \bibinfo {author} {\bibfnamefont {Y.}~\bibnamefont
  {Yi}}, \bibinfo {author} {\bibfnamefont {Y.~Z.}\ \bibnamefont {Wu}}, \bibinfo
  {author} {\bibfnamefont {S.}~\bibnamefont {Wu}}, \bibinfo {author}
  {\bibfnamefont {J.}~\bibnamefont {Zhu}}, \bibinfo {author} {\bibfnamefont
  {J.}~\bibnamefont {Wang}}, \bibinfo {author} {\bibfnamefont {X.~H.}\
  \bibnamefont {Chen}},\ and\ \bibinfo {author} {\bibfnamefont
  {Y.}~\bibnamefont {Zhang}},\ }\href
  {https://www.nature.com/articles/s41586-018-0626-9} {\bibfield  {journal}
  {\bibinfo  {journal} {Nature}\ }\textbf {\bibinfo {volume} {563}},\ \bibinfo
  {pages} {94} (\bibinfo {year} {2018})}\BibitemShut {NoStop}%
\bibitem [{\citenamefont {Fei}\ \emph {et~al.}(2018)\citenamefont {Fei},
  \citenamefont {Huang}, \citenamefont {Malinowski}, \citenamefont {Wang},
  \citenamefont {Song}, \citenamefont {Sanchez}, \citenamefont {Yao},
  \citenamefont {Xiao}, \citenamefont {Zhu}, \citenamefont {May}, \citenamefont
  {Wu}, \citenamefont {Cobden}, \citenamefont {Chu},\ and\ \citenamefont
  {Xu}}]{Fei2018}%
  \BibitemOpen
  \bibfield  {author} {\bibinfo {author} {\bibfnamefont {Z.}~\bibnamefont
  {Fei}}, \bibinfo {author} {\bibfnamefont {B.}~\bibnamefont {Huang}}, \bibinfo
  {author} {\bibfnamefont {P.}~\bibnamefont {Malinowski}}, \bibinfo {author}
  {\bibfnamefont {W.}~\bibnamefont {Wang}}, \bibinfo {author} {\bibfnamefont
  {T.}~\bibnamefont {Song}}, \bibinfo {author} {\bibfnamefont {J.}~\bibnamefont
  {Sanchez}}, \bibinfo {author} {\bibfnamefont {W.}~\bibnamefont {Yao}},
  \bibinfo {author} {\bibfnamefont {D.}~\bibnamefont {Xiao}}, \bibinfo {author}
  {\bibfnamefont {X.}~\bibnamefont {Zhu}}, \bibinfo {author} {\bibfnamefont
  {A.~F.}\ \bibnamefont {May}}, \bibinfo {author} {\bibfnamefont
  {W.}~\bibnamefont {Wu}}, \bibinfo {author} {\bibfnamefont {D.~H.}\
  \bibnamefont {Cobden}}, \bibinfo {author} {\bibfnamefont {J.-H.}\
  \bibnamefont {Chu}},\ and\ \bibinfo {author} {\bibfnamefont {X.}~\bibnamefont
  {Xu}},\ }\href {https://www.nature.com/articles/s41563-018-0149-7} {\bibfield
   {journal} {\bibinfo  {journal} {Nat. Mater.}\ }\textbf {\bibinfo {volume}
  {17}},\ \bibinfo {pages} {778} (\bibinfo {year} {2018})}\BibitemShut
  {NoStop}%
\bibitem [{\citenamefont {May}\ \emph {et~al.}(2016)\citenamefont {May},
  \citenamefont {Calder}, \citenamefont {Cantoni}, \citenamefont {Cao},\ and\
  \citenamefont {McGuire}}]{May2016}%
  \BibitemOpen
  \bibfield  {author} {\bibinfo {author} {\bibfnamefont {A.~F.}\ \bibnamefont
  {May}}, \bibinfo {author} {\bibfnamefont {S.}~\bibnamefont {Calder}},
  \bibinfo {author} {\bibfnamefont {C.}~\bibnamefont {Cantoni}}, \bibinfo
  {author} {\bibfnamefont {H.}~\bibnamefont {Cao}},\ and\ \bibinfo {author}
  {\bibfnamefont {M.~A.}\ \bibnamefont {McGuire}},\ }\href
  {https://journals.aps.org/prb/abstract/10.1103/PhysRevB.93.014411} {\bibfield
   {journal} {\bibinfo  {journal} {Phys. Rev. B.}\ }\textbf {\bibinfo {volume}
  {93}},\ \bibinfo {pages} {014411} (\bibinfo {year} {2016})}\BibitemShut
  {NoStop}%
\bibitem [{\citenamefont {May}\ \emph {et~al.}(2019)\citenamefont {May},
  \citenamefont {Ovchinnikov}, \citenamefont {Zheng}, \citenamefont {Hermann},
  \citenamefont {Calder}, \citenamefont {Huang}, \citenamefont {Fei},
  \citenamefont {Liu}, \citenamefont {Xu},\ and\ \citenamefont
  {McGuire}}]{May2019}%
  \BibitemOpen
  \bibfield  {author} {\bibinfo {author} {\bibfnamefont {A.~F.}\ \bibnamefont
  {May}}, \bibinfo {author} {\bibfnamefont {D.}~\bibnamefont {Ovchinnikov}},
  \bibinfo {author} {\bibfnamefont {Q.}~\bibnamefont {Zheng}}, \bibinfo
  {author} {\bibfnamefont {R.}~\bibnamefont {Hermann}}, \bibinfo {author}
  {\bibfnamefont {S.}~\bibnamefont {Calder}}, \bibinfo {author} {\bibfnamefont
  {B.}~\bibnamefont {Huang}}, \bibinfo {author} {\bibfnamefont
  {Z.}~\bibnamefont {Fei}}, \bibinfo {author} {\bibfnamefont {Y.}~\bibnamefont
  {Liu}}, \bibinfo {author} {\bibfnamefont {X.}~\bibnamefont {Xu}},\ and\
  \bibinfo {author} {\bibfnamefont {M.~A.}\ \bibnamefont {McGuire}},\ }\href
  {https://pubs.acs.org/doi/10.1021/acsnano.8b09660} {\bibfield  {journal}
  {\bibinfo  {journal} {ACS Nano}\ }\textbf {\bibinfo {volume} {13}},\ \bibinfo
  {pages} {4436} (\bibinfo {year} {2019})}\BibitemShut {NoStop}%
\bibitem [{\citenamefont {Nakano}\ \emph {et~al.}(2019)\citenamefont {Nakano},
  \citenamefont {Wang}, \citenamefont {Yoshida}, \citenamefont {Matsuoka},
  \citenamefont {Majima}, \citenamefont {Ikeda}, \citenamefont {Hirata},
  \citenamefont {Takeda}, \citenamefont {Wadati}, \citenamefont {Kohama},
  \citenamefont {Ohigashi}, \citenamefont {Sakano}, \citenamefont {Ishizaka},\
  and\ \citenamefont {Iwasa}}]{Nakano2019}%
  \BibitemOpen
  \bibfield  {author} {\bibinfo {author} {\bibfnamefont {M.}~\bibnamefont
  {Nakano}}, \bibinfo {author} {\bibfnamefont {Y.}~\bibnamefont {Wang}},
  \bibinfo {author} {\bibfnamefont {S.}~\bibnamefont {Yoshida}}, \bibinfo
  {author} {\bibfnamefont {H.}~\bibnamefont {Matsuoka}}, \bibinfo {author}
  {\bibfnamefont {Y.}~\bibnamefont {Majima}}, \bibinfo {author} {\bibfnamefont
  {K.}~\bibnamefont {Ikeda}}, \bibinfo {author} {\bibfnamefont
  {Y.}~\bibnamefont {Hirata}}, \bibinfo {author} {\bibfnamefont
  {Y.}~\bibnamefont {Takeda}}, \bibinfo {author} {\bibfnamefont
  {H.}~\bibnamefont {Wadati}}, \bibinfo {author} {\bibfnamefont
  {Y.}~\bibnamefont {Kohama}}, \bibinfo {author} {\bibfnamefont
  {Y.}~\bibnamefont {Ohigashi}}, \bibinfo {author} {\bibfnamefont
  {M.}~\bibnamefont {Sakano}}, \bibinfo {author} {\bibfnamefont
  {K.}~\bibnamefont {Ishizaka}},\ and\ \bibinfo {author} {\bibfnamefont
  {Y.}~\bibnamefont {Iwasa}},\ }\href
  {https://doi.org/10.1021/acs.nanolett.9b03614} {\bibfield  {journal}
  {\bibinfo  {journal} {Nano Letters}\ }\textbf {\bibinfo {volume} {19}},\
  \bibinfo {pages} {8806} (\bibinfo {year} {2019})}\BibitemShut {NoStop}%
\bibitem [{\citenamefont {Seo}\ \emph {et~al.}(2020)\citenamefont {Seo},
  \citenamefont {Kim}, \citenamefont {An}, \citenamefont {Kim}, \citenamefont
  {Kim}, \citenamefont {Hwang}, \citenamefont {Kim}, \citenamefont {Jang},
  \citenamefont {Kim}, \citenamefont {Eom}, \citenamefont {Seo}, \citenamefont
  {Stania}, \citenamefont {Muntwiler}, \citenamefont {Lee}, \citenamefont
  {Watanabe}, \citenamefont {Taniguchi}, \citenamefont {Jo}, \citenamefont
  {Lee}, \citenamefont {Min}, \citenamefont {Jo}, \citenamefont {Yeom},
  \citenamefont {Choi}, \citenamefont {Shim},\ and\ \citenamefont
  {Kim}}]{Seo2020}%
  \BibitemOpen
  \bibfield  {author} {\bibinfo {author} {\bibfnamefont {J.}~\bibnamefont
  {Seo}}, \bibinfo {author} {\bibfnamefont {D.~Y.}\ \bibnamefont {Kim}},
  \bibinfo {author} {\bibfnamefont {E.~S.}\ \bibnamefont {An}}, \bibinfo
  {author} {\bibfnamefont {K.}~\bibnamefont {Kim}}, \bibinfo {author}
  {\bibfnamefont {G.-Y.}\ \bibnamefont {Kim}}, \bibinfo {author} {\bibfnamefont
  {S.-Y.}\ \bibnamefont {Hwang}}, \bibinfo {author} {\bibfnamefont {D.~W.}\
  \bibnamefont {Kim}}, \bibinfo {author} {\bibfnamefont {B.~G.}\ \bibnamefont
  {Jang}}, \bibinfo {author} {\bibfnamefont {H.}~\bibnamefont {Kim}}, \bibinfo
  {author} {\bibfnamefont {G.}~\bibnamefont {Eom}}, \bibinfo {author}
  {\bibfnamefont {S.~Y.}\ \bibnamefont {Seo}}, \bibinfo {author} {\bibfnamefont
  {R.}~\bibnamefont {Stania}}, \bibinfo {author} {\bibfnamefont
  {M.}~\bibnamefont {Muntwiler}}, \bibinfo {author} {\bibfnamefont
  {J.}~\bibnamefont {Lee}}, \bibinfo {author} {\bibfnamefont {K.}~\bibnamefont
  {Watanabe}}, \bibinfo {author} {\bibfnamefont {T.}~\bibnamefont {Taniguchi}},
  \bibinfo {author} {\bibfnamefont {Y.~J.}\ \bibnamefont {Jo}}, \bibinfo
  {author} {\bibfnamefont {J.}~\bibnamefont {Lee}}, \bibinfo {author}
  {\bibfnamefont {B.~I.}\ \bibnamefont {Min}}, \bibinfo {author} {\bibfnamefont
  {M.~H.}\ \bibnamefont {Jo}}, \bibinfo {author} {\bibfnamefont {H.~W.}\
  \bibnamefont {Yeom}}, \bibinfo {author} {\bibfnamefont {S.-Y.}\ \bibnamefont
  {Choi}}, \bibinfo {author} {\bibfnamefont {J.~H.}\ \bibnamefont {Shim}},\
  and\ \bibinfo {author} {\bibfnamefont {J.~S.}\ \bibnamefont {Kim}},\ }\href
  {https://www.science.org/doi/10.1126/sciadv.aay8912} {\bibfield  {journal}
  {\bibinfo  {journal} {Sci. Adv.}\ }\textbf {\bibinfo {volume} {6}},\ \bibinfo
  {pages} {eaay8912} (\bibinfo {year} {2020})}\BibitemShut {NoStop}%
\bibitem [{\citenamefont {Hubbard}(1963)}]{Hubbard1963}%
  \BibitemOpen
  \bibfield  {author} {\bibinfo {author} {\bibfnamefont {J.}~\bibnamefont
  {Hubbard}},\ }\href
  {https://royalsocietypublishing.org/doi/10.1098/rspa.1963.0204} {\bibfield
  {journal} {\bibinfo  {journal} {Proc. R. Soc. Lond.}\ }\textbf {\bibinfo
  {volume} {276}},\ \bibinfo {pages} {238} (\bibinfo {year}
  {1963})}\BibitemShut {NoStop}%
\bibitem [{\citenamefont {Gutzwiller}(1963)}]{Gutzwiller1963}%
  \BibitemOpen
  \bibfield  {author} {\bibinfo {author} {\bibfnamefont {M.~C.}\ \bibnamefont
  {Gutzwiller}},\ }\href
  {https://journals.aps.org/prl/abstract/10.1103/PhysRevLett.10.159} {\bibfield
   {journal} {\bibinfo  {journal} {Phys. Rev. Lett.}\ }\textbf {\bibinfo
  {volume} {10}},\ \bibinfo {pages} {159} (\bibinfo {year} {1963})}\BibitemShut
  {NoStop}%
\bibitem [{\citenamefont {Kanamori}(1963)}]{Kanamori1963}%
  \BibitemOpen
  \bibfield  {author} {\bibinfo {author} {\bibfnamefont {J.}~\bibnamefont
  {Kanamori}},\ }\href {https://academic.oup.com/ptp/article/30/3/275/1865799}
  {\bibfield  {journal} {\bibinfo  {journal} {Progr. Theoret. Phys.}\ }\textbf
  {\bibinfo {volume} {30}},\ \bibinfo {pages} {275} (\bibinfo {year}
  {1963})}\BibitemShut {NoStop}%
\bibitem [{\citenamefont {Nagaoka}(1966)}]{Nagaoka}%
  \BibitemOpen
  \bibfield  {author} {\bibinfo {author} {\bibfnamefont {Y.}~\bibnamefont
  {Nagaoka}},\ }\href {https://doi.org/10.1103/PhysRev.147.392} {\bibfield
  {journal} {\bibinfo  {journal} {Phys. Rev.}\ }\textbf {\bibinfo {volume}
  {147}},\ \bibinfo {pages} {392} (\bibinfo {year} {1966})}\BibitemShut
  {NoStop}%
\bibitem [{\citenamefont {Moriya}\ and\ \citenamefont
  {Kawabata}(1973)}]{Moriya1973}%
  \BibitemOpen
  \bibfield  {author} {\bibinfo {author} {\bibfnamefont {T.}~\bibnamefont
  {Moriya}}\ and\ \bibinfo {author} {\bibfnamefont {A.}~\bibnamefont
  {Kawabata}},\ }\href {https://journals.jps.jp/doi/10.1143/JPSJ.34.639}
  {\bibfield  {journal} {\bibinfo  {journal} {J. Phys. Soc. Jpn.}\ }\textbf
  {\bibinfo {volume} {34}},\ \bibinfo {pages} {639} (\bibinfo {year}
  {1973})}\BibitemShut {NoStop}%
\bibitem [{\citenamefont {Tasaki}(1998)}]{Tasaki1998}%
  \BibitemOpen
  \bibfield  {author} {\bibinfo {author} {\bibfnamefont {H.}~\bibnamefont
  {Tasaki}},\ }\href {https://dx.doi.org/10.1143/PTP.99.489} {\bibfield
  {journal} {\bibinfo  {journal} {Prog. Theor. Phys.}\ }\textbf {\bibinfo
  {volume} {99}},\ \bibinfo {pages} {489} (\bibinfo {year} {1998})}\BibitemShut
  {NoStop}%
\bibitem [{\citenamefont {Stoner}(1936)}]{Stoner1936}%
  \BibitemOpen
  \bibfield  {author} {\bibinfo {author} {\bibfnamefont {E.~C.}\ \bibnamefont
  {Stoner}},\ }\href {https://doi.org/10.1098/rspa.1936.0075} {\bibfield
  {journal} {\bibinfo  {journal} {Proceedings of the Royal Society of London.
  A. Mathematical and Physical Sciences}\ }\textbf {\bibinfo {volume} {154}},\
  \bibinfo {pages} {656} (\bibinfo {year} {1936})},\ \Eprint
  {https://arxiv.org/abs/https://royalsocietypublishing.org/rspa/article-pdf/154/883/656/33982/rspa.1936.0075.pdf}
  {https://royalsocietypublishing.org/rspa/article-pdf/154/883/656/33982/rspa.1936.0075.pdf}
  \BibitemShut {NoStop}%
\bibitem [{\citenamefont {Mielke}\ and\ \citenamefont
  {Tasaki}(1993)}]{Mielke1993}%
  \BibitemOpen
  \bibfield  {author} {\bibinfo {author} {\bibfnamefont {A.}~\bibnamefont
  {Mielke}}\ and\ \bibinfo {author} {\bibfnamefont {H.}~\bibnamefont
  {Tasaki}},\ }\href {https://link.springer.com/article/10.1007/BF02108079}
  {\bibfield  {journal} {\bibinfo  {journal} {Commun. Math. Phys.}\ }\textbf
  {\bibinfo {volume} {158}},\ \bibinfo {pages} {341} (\bibinfo {year}
  {1993})}\BibitemShut {NoStop}%
\bibitem [{\citenamefont {Mielke}(1999)}]{Mielke1999}%
  \BibitemOpen
  \bibfield  {author} {\bibinfo {author} {\bibfnamefont {A.}~\bibnamefont
  {Mielke}},\ }\href
  {https://iopscience.iop.org/article/10.1088/0305-4470/32/48/304} {\bibfield
  {journal} {\bibinfo  {journal} {J. Phys. A Math. Gen.}\ }\textbf {\bibinfo
  {volume} {32}},\ \bibinfo {pages} {8411} (\bibinfo {year}
  {1999})}\BibitemShut {NoStop}%
\bibitem [{\citenamefont {Katsura}\ \emph {et~al.}(2010)\citenamefont
  {Katsura}, \citenamefont {Maruyama}, \citenamefont {Tanaka},\ and\
  \citenamefont {Tasaki}}]{Katsura2010}%
  \BibitemOpen
  \bibfield  {author} {\bibinfo {author} {\bibfnamefont {H.}~\bibnamefont
  {Katsura}}, \bibinfo {author} {\bibfnamefont {I.}~\bibnamefont {Maruyama}},
  \bibinfo {author} {\bibfnamefont {A.}~\bibnamefont {Tanaka}},\ and\ \bibinfo
  {author} {\bibfnamefont {H.}~\bibnamefont {Tasaki}},\ }\href
  {https://iopscience.iop.org/article/10.1209/0295-5075/91/57007} {\bibfield
  {journal} {\bibinfo  {journal} {EPL}\ }\textbf {\bibinfo {volume} {91}},\
  \bibinfo {pages} {57007} (\bibinfo {year} {2010})}\BibitemShut {NoStop}%
\bibitem [{\citenamefont {Tasaki}()}]{Tasaki2020}%
  \BibitemOpen
  \bibfield  {author} {\bibinfo {author} {\bibfnamefont {H.}~\bibnamefont
  {Tasaki}},\ }\href
  {https://doi.org/https://doi.org/10.1007/978-3-030-41265-4} {\bibinfo {title}
  {Physics and mathematics of quantum many-body systems}},\ \bibinfo {note}
  {(Springer, \TK{Cham}, 2020)}\BibitemShut {NoStop}%
\bibitem [{\citenamefont {Lieb}(1989)}]{Lieb1989}%
  \BibitemOpen
  \bibfield  {author} {\bibinfo {author} {\bibfnamefont {E.~H.}\ \bibnamefont
  {Lieb}},\ }\href {https://doi.org/10.1103/PhysRevLett.62.1201} {\bibfield
  {journal} {\bibinfo  {journal} {Phys. Rev. Lett.}\ }\textbf {\bibinfo
  {volume} {62}},\ \bibinfo {pages} {1201} (\bibinfo {year}
  {1989})}\BibitemShut {NoStop}%
\bibitem [{\citenamefont {Mielke}(1991)}]{Mielke1991}%
  \BibitemOpen
  \bibfield  {author} {\bibinfo {author} {\bibfnamefont {A.}~\bibnamefont
  {Mielke}},\ }\href
  {https://iopscience.iop.org/article/10.1088/0305-4470/24/14/018/meta?casa_token=vm7VoUl8QqAAAAAA:iRMZNE4jfTxpcUAn-zTKVYhpqDgOZ491uKS_pUY0r9Xxv1rNsrvqPogsRHMdAoKuyg09nkjDG4mFMP0gdj2-1zHUbKAFCQ}
  {\bibfield  {journal} {\bibinfo  {journal} {J. Phys. A Math. Gen.}\ }\textbf
  {\bibinfo {volume} {24}},\ \bibinfo {pages} {3311} (\bibinfo {year}
  {1991})}\BibitemShut {NoStop}%
\bibitem [{\citenamefont {Mielke}(1992)}]{Mielke1992}%
  \BibitemOpen
  \bibfield  {author} {\bibinfo {author} {\bibfnamefont {A.}~\bibnamefont
  {Mielke}},\ }\href
  {https://iopscience.iop.org/article/10.1088/0305-4470/25/16/011} {\bibfield
  {journal} {\bibinfo  {journal} {J. Phys. A Math. Gen.}\ }\textbf {\bibinfo
  {volume} {25}},\ \bibinfo {pages} {4335} (\bibinfo {year}
  {1992})}\BibitemShut {NoStop}%
\bibitem [{\citenamefont {Tasaki}(1992)}]{Tasaki1992}%
  \BibitemOpen
  \bibfield  {author} {\bibinfo {author} {\bibfnamefont {H.}~\bibnamefont
  {Tasaki}},\ }\href
  {https://journals.aps.org/prl/abstract/10.1103/PhysRevLett.69.1608}
  {\bibfield  {journal} {\bibinfo  {journal} {Phys. Rev. Lett.}\ }\textbf
  {\bibinfo {volume} {69}},\ \bibinfo {pages} {1608} (\bibinfo {year}
  {1992})}\BibitemShut {NoStop}%
\bibitem [{\citenamefont {Tasaki}(1995)}]{Tasaki1995}%
  \BibitemOpen
  \bibfield  {author} {\bibinfo {author} {\bibfnamefont {H.}~\bibnamefont
  {Tasaki}},\ }\href
  {https://journals.aps.org/prl/abstract/10.1103/PhysRevLett.75.4678}
  {\bibfield  {journal} {\bibinfo  {journal} {Phys. Rev. Lett.}\ }\textbf
  {\bibinfo {volume} {75}},\ \bibinfo {pages} {4678} (\bibinfo {year}
  {1995})}\BibitemShut {NoStop}%
\bibitem [{\citenamefont {Tanaka}\ and\ \citenamefont
  {Idogaki}(2001)}]{Tanaka2001}%
  \BibitemOpen
  \bibfield  {author} {\bibinfo {author} {\bibfnamefont {A.}~\bibnamefont
  {Tanaka}}\ and\ \bibinfo {author} {\bibfnamefont {T.}~\bibnamefont
  {Idogaki}},\ }\href
  {https://www.sciencedirect.com/science/article/abs/pii/S0378437101002278?via%3Dihub}
  {\bibfield  {journal} {\bibinfo  {journal} {Physica A}\ }\textbf {\bibinfo
  {volume} {297}},\ \bibinfo {pages} {441} (\bibinfo {year}
  {2001})}\BibitemShut {NoStop}%
\bibitem [{\citenamefont {Tanaka}\ and\ \citenamefont
  {Ueda}(2003)}]{Tanaka2003}%
  \BibitemOpen
  \bibfield  {author} {\bibinfo {author} {\bibfnamefont {A.}~\bibnamefont
  {Tanaka}}\ and\ \bibinfo {author} {\bibfnamefont {H.}~\bibnamefont {Ueda}},\
  }\href {https://journals.aps.org/prl/abstract/10.1103/PhysRevLett.90.067204}
  {\bibfield  {journal} {\bibinfo  {journal} {Phys. Rev. Lett.}\ }\textbf
  {\bibinfo {volume} {90}},\ \bibinfo {pages} {067204} (\bibinfo {year}
  {2003})}\BibitemShut {NoStop}%
\bibitem [{\citenamefont {Ueda}\ \emph {et~al.}(2004)\citenamefont {Ueda},
  \citenamefont {Tanaka},\ and\ \citenamefont {Idogaki}}]{Ueda2004}%
  \BibitemOpen
  \bibfield  {author} {\bibinfo {author} {\bibfnamefont {H.}~\bibnamefont
  {Ueda}}, \bibinfo {author} {\bibfnamefont {A.}~\bibnamefont {Tanaka}},\ and\
  \bibinfo {author} {\bibfnamefont {T.}~\bibnamefont {Idogaki}},\ }\href
  {https://www.sciencedirect.com/science/article/abs/pii/S0304885303026283?via%3Dihub}
  {\bibfield  {journal} {\bibinfo  {journal} {J. Magn. Magn. Mater.}\ }\textbf
  {\bibinfo {volume} {272-276}},\ \bibinfo {pages} {950} (\bibinfo {year}
  {2004})}\BibitemShut {NoStop}%
\bibitem [{\citenamefont {Lu}(2009)}]{Lu2009}%
  \BibitemOpen
  \bibfield  {author} {\bibinfo {author} {\bibfnamefont {L.}~\bibnamefont
  {Lu}},\ }\href
  {https://iopscience.iop.org/article/10.1088/1751-8113/42/26/265002}
  {\bibfield  {journal} {\bibinfo  {journal} {J. Phys. A Math. Theor.}\
  }\textbf {\bibinfo {volume} {42}},\ \bibinfo {pages} {265002} (\bibinfo
  {year} {2009})}\BibitemShut {NoStop}%
\bibitem [{\citenamefont {Tanaka}(2018)}]{Tanaka2018}%
  \BibitemOpen
  \bibfield  {author} {\bibinfo {author} {\bibfnamefont {A.}~\bibnamefont
  {Tanaka}},\ }\href
  {https://link.springer.com/article/10.1007/s10955-017-1932-6} {\bibfield
  {journal} {\bibinfo  {journal} {J. Stat. Phys.}\ }\textbf {\bibinfo {volume}
  {170}},\ \bibinfo {pages} {399} (\bibinfo {year} {2018})}\BibitemShut
  {NoStop}%
\bibitem [{\citenamefont {Tamura}\ and\ \citenamefont
  {Katsura}(2019)}]{Tamura2019}%
  \BibitemOpen
  \bibfield  {author} {\bibinfo {author} {\bibfnamefont {K.}~\bibnamefont
  {Tamura}}\ and\ \bibinfo {author} {\bibfnamefont {H.}~\bibnamefont
  {Katsura}},\ }\href
  {https://journals.aps.org/prb/abstract/10.1103/PhysRevB.100.214423}
  {\bibfield  {journal} {\bibinfo  {journal} {Phys. Rev. B.}\ }\textbf
  {\bibinfo {volume} {100}},\ \bibinfo {pages} {214423} (\bibinfo {year}
  {2019})}\BibitemShut {NoStop}%
\bibitem [{\citenamefont {Honerkamp}\ and\ \citenamefont
  {Salmhofer}(2001)}]{Honerkamp2001}%
  \BibitemOpen
  \bibfield  {author} {\bibinfo {author} {\bibfnamefont {C.}~\bibnamefont
  {Honerkamp}}\ and\ \bibinfo {author} {\bibfnamefont {M.}~\bibnamefont
  {Salmhofer}},\ }\href
  {https://journals.aps.org/prl/abstract/10.1103/PhysRevLett.87.187004}
  {\bibfield  {journal} {\bibinfo  {journal} {Phys. Rev. Lett.}\ }\textbf
  {\bibinfo {volume} {87}},\ \bibinfo {pages} {187004} (\bibinfo {year}
  {2001})}\BibitemShut {NoStop}%
\bibitem [{\citenamefont {Yu.~Irkhin}\ \emph {et~al.}(2001)\citenamefont
  {Yu.~Irkhin}, \citenamefont {Katanin},\ and\ \citenamefont
  {Katsnelson}}]{Yu-Irkhin2001}%
  \BibitemOpen
  \bibfield  {author} {\bibinfo {author} {\bibfnamefont {V.}~\bibnamefont
  {Yu.~Irkhin}}, \bibinfo {author} {\bibfnamefont {A.~A.}\ \bibnamefont
  {Katanin}},\ and\ \bibinfo {author} {\bibfnamefont {M.~I.}\ \bibnamefont
  {Katsnelson}},\ }\href
  {https://journals-aps-org.kyoto-u.idm.oclc.org/prb/abstract/10.1103/PhysRevB.64.165107}
  {\bibfield  {journal} {\bibinfo  {journal} {Phys. Rev. B}\ }\textbf {\bibinfo
  {volume} {64}},\ \bibinfo {pages} {165107} (\bibinfo {year}
  {2001})}\BibitemShut {NoStop}%
\bibitem [{\citenamefont {Katanin}\ and\ \citenamefont
  {Kampf}(2003)}]{Katanin2003}%
  \BibitemOpen
  \bibfield  {author} {\bibinfo {author} {\bibfnamefont {A.~A.}\ \bibnamefont
  {Katanin}}\ and\ \bibinfo {author} {\bibfnamefont {A.~P.}\ \bibnamefont
  {Kampf}},\ }\href
  {https://journals-aps-org.kyoto-u.idm.oclc.org/prb/abstract/10.1103/PhysRevB.68.195101}
  {\bibfield  {journal} {\bibinfo  {journal} {Phys. Rev. B}\ }\textbf {\bibinfo
  {volume} {68}},\ \bibinfo {pages} {195101} (\bibinfo {year}
  {2003})}\BibitemShut {NoStop}%
\bibitem [{\citenamefont {Yuan}\ \emph {et~al.}(2019)\citenamefont {Yuan},
  \citenamefont {Isobe},\ and\ \citenamefont {Fu}}]{Yuan2019-hm}%
  \BibitemOpen
  \bibfield  {author} {\bibinfo {author} {\bibfnamefont {N.~F.~Q.}\
  \bibnamefont {Yuan}}, \bibinfo {author} {\bibfnamefont {H.}~\bibnamefont
  {Isobe}},\ and\ \bibinfo {author} {\bibfnamefont {L.}~\bibnamefont {Fu}},\
  }\href {https://www.nature.com/articles/s41467-019-13670-9} {\bibfield
  {journal} {\bibinfo  {journal} {Nat. Commun.}\ }\textbf {\bibinfo {volume}
  {10}},\ \bibinfo {pages} {5769} (\bibinfo {year} {2019})}\BibitemShut
  {NoStop}%
\bibitem [{\citenamefont {Classen}\ and\ \citenamefont
  {Betouras}(2025)}]{Classen2025-kl}%
  \BibitemOpen
  \bibfield  {author} {\bibinfo {author} {\bibfnamefont {L.}~\bibnamefont
  {Classen}}\ and\ \bibinfo {author} {\bibfnamefont {J.~J.}\ \bibnamefont
  {Betouras}},\ }\href
  {https://www.annualreviews.org/content/journals/10.1146/annurev-conmatphys-042924-015000}
  {\bibfield  {journal} {\bibinfo  {journal} {Annu. Rev. Condens. Matter
  Phys.}\ }\textbf {\bibinfo {volume} {16}},\ \bibinfo {pages} {229} (\bibinfo
  {year} {2025})}\BibitemShut {NoStop}%
\bibitem [{\citenamefont {Yanase}\ \emph {et~al.}(2003)\citenamefont {Yanase},
  \citenamefont {Jujo}, \citenamefont {Nomura}, \citenamefont {Ikeda},
  \citenamefont {Hotta},\ and\ \citenamefont {Yamada}}]{Yanase2003}%
  \BibitemOpen
  \bibfield  {author} {\bibinfo {author} {\bibfnamefont {Y.}~\bibnamefont
  {Yanase}}, \bibinfo {author} {\bibfnamefont {T.}~\bibnamefont {Jujo}},
  \bibinfo {author} {\bibfnamefont {T.}~\bibnamefont {Nomura}}, \bibinfo
  {author} {\bibfnamefont {H.}~\bibnamefont {Ikeda}}, \bibinfo {author}
  {\bibfnamefont {T.}~\bibnamefont {Hotta}},\ and\ \bibinfo {author}
  {\bibfnamefont {K.}~\bibnamefont {Yamada}},\ }\href
  {https://www.sciencedirect.com/science/article/abs/pii/S0370157303003235}
  {\bibfield  {journal} {\bibinfo  {journal} {Phys. Rep.}\ }\textbf {\bibinfo
  {volume} {387}},\ \bibinfo {pages} {1} (\bibinfo {year} {2003})}\BibitemShut
  {NoStop}%
\bibitem [{\citenamefont {Kitamura}\ \emph {et~al.}(2024)\citenamefont
  {Kitamura}, \citenamefont {Daido},\ and\ \citenamefont
  {Yanase}}]{Kitamura2024}%
  \BibitemOpen
  \bibfield  {author} {\bibinfo {author} {\bibfnamefont {T.}~\bibnamefont
  {Kitamura}}, \bibinfo {author} {\bibfnamefont {A.}~\bibnamefont {Daido}},\
  and\ \bibinfo {author} {\bibfnamefont {Y.}~\bibnamefont {Yanase}},\ }\href
  {http://dx.doi.org/10.1103/PhysRevLett.132.036001} {\bibfield  {journal}
  {\bibinfo  {journal} {Phys. Rev. Lett.}\ }\textbf {\bibinfo {volume} {132}},\
  \bibinfo {pages} {036001} (\bibinfo {year} {2024})}\BibitemShut {NoStop}%
\bibitem [{\citenamefont {Resta}(2011)}]{Resta2011}%
  \BibitemOpen
  \bibfield  {author} {\bibinfo {author} {\bibfnamefont {R.}~\bibnamefont
  {Resta}},\ }\href
  {https://link.springer.com/article/10.1140/epjb/e2010-10874-4} {\bibfield
  {journal} {\bibinfo  {journal} {Eur. Phys. J. B}\ }\textbf {\bibinfo {volume}
  {79}},\ \bibinfo {pages} {121} (\bibinfo {year} {2011})}\BibitemShut
  {NoStop}%
\bibitem [{\citenamefont {Cheng}(2013)}]{Cheng2013}%
  \BibitemOpen
  \bibfield  {author} {\bibinfo {author} {\bibfnamefont {R.}~\bibnamefont
  {Cheng}},\ }\href {https://arxiv.org/abs/1012.1337} {\bibinfo {title}
  {Quantum geometric tensor (fubini-study metric) in simple quantum system: A
  pedagogical introduction}} (\bibinfo {year} {2013}),\ \Eprint
  {https://arxiv.org/abs/1012.1337} {arXiv:1012.1337 [quant-ph]} \BibitemShut
  {NoStop}%
\bibitem [{\citenamefont {Rossi}(2021)}]{Rossi2021}%
  \BibitemOpen
  \bibfield  {author} {\bibinfo {author} {\bibfnamefont {E.}~\bibnamefont
  {Rossi}},\ }\href
  {https://www.sciencedirect.com/science/article/abs/pii/S1359028621000553?via%3Dihub}
  {\bibfield  {journal} {\bibinfo  {journal} {Curr. Opin. Solid State Mater.
  Sci.}\ }\textbf {\bibinfo {volume} {25}},\ \bibinfo {pages} {100952}
  (\bibinfo {year} {2021})}\BibitemShut {NoStop}%
\bibitem [{\citenamefont {Törmä}\ \emph {et~al.}(2022)\citenamefont
  {Törmä}, \citenamefont {Peotta},\ and\ \citenamefont
  {Bernevig}}]{Torma2022}%
  \BibitemOpen
  \bibfield  {author} {\bibinfo {author} {\bibfnamefont {P.}~\bibnamefont
  {Törmä}}, \bibinfo {author} {\bibfnamefont {S.}~\bibnamefont {Peotta}},\
  and\ \bibinfo {author} {\bibfnamefont {B.~A.}\ \bibnamefont {Bernevig}},\
  }\href {http://dx.doi.org/10.1038/s42254-022-00466-y} {\bibfield  {journal}
  {\bibinfo  {journal} {Nature Reviews Physics}\ }\textbf {\bibinfo {volume}
  {4}},\ \bibinfo {pages} {528} (\bibinfo {year} {2022})}\BibitemShut {NoStop}%
\bibitem [{\citenamefont {Törmä}(2023)}]{Torma2023}%
  \BibitemOpen
  \bibfield  {author} {\bibinfo {author} {\bibfnamefont {P.}~\bibnamefont
  {Törmä}},\ }\href
  {https://journals.aps.org/prl/abstract/10.1103/PhysRevLett.131.240001}
  {\bibfield  {journal} {\bibinfo  {journal} {Phys. Rev. Lett.}\ }\textbf
  {\bibinfo {volume} {131}},\ \bibinfo {pages} {240001} (\bibinfo {year}
  {2023})}\BibitemShut {NoStop}%
\bibitem [{\citenamefont {Yu}\ \emph {et~al.}(2025)\citenamefont {Yu},
  \citenamefont {Bernevig}, \citenamefont {Queiroz}, \citenamefont {Rossi},
  \citenamefont {Törmä},\ and\ \citenamefont {Yang}}]{Yu2024}%
  \BibitemOpen
  \bibfield  {author} {\bibinfo {author} {\bibfnamefont {J.}~\bibnamefont
  {Yu}}, \bibinfo {author} {\bibfnamefont {B.~A.}\ \bibnamefont {Bernevig}},
  \bibinfo {author} {\bibfnamefont {R.}~\bibnamefont {Queiroz}}, \bibinfo
  {author} {\bibfnamefont {E.}~\bibnamefont {Rossi}}, \bibinfo {author}
  {\bibfnamefont {P.}~\bibnamefont {Törmä}},\ and\ \bibinfo {author}
  {\bibfnamefont {B.-J.}\ \bibnamefont {Yang}},\ }\href
  {https://www.nature.com/articles/s41535-025-00801-3} {\bibfield  {journal}
  {\bibinfo  {journal} {Npj Quantum Mater.}\ }\textbf {\bibinfo {volume}
  {10}},\ \bibinfo {pages} {\TK{101}} (\bibinfo {year} {2025})}\BibitemShut
  {NoStop}%
\bibitem [{\citenamefont {Herzog-Arbeitman}\ \emph {et~al.}(2026)\citenamefont
  {Herzog-Arbeitman}, \citenamefont {Chew}, \citenamefont {Huhtinen},
  \citenamefont {Törmä},\ and\ \citenamefont {Bernevig}}]{Herzog2022}%
  \BibitemOpen
  \bibfield  {author} {\bibinfo {author} {\bibfnamefont {J.}~\bibnamefont
  {Herzog-Arbeitman}}, \bibinfo {author} {\bibfnamefont {A.}~\bibnamefont
  {Chew}}, \bibinfo {author} {\bibfnamefont {K.-E.}\ \bibnamefont {Huhtinen}},
  \bibinfo {author} {\bibfnamefont {P.}~\bibnamefont {Törmä}},\ and\ \bibinfo
  {author} {\bibfnamefont {B.~A.}\ \bibnamefont {Bernevig}},\ }\href
  {https://www.nature.com/articles/s42005-026-02732-2} {\bibfield  {journal}
  {\bibinfo  {journal} {Commun. Phys.}\ }\textbf {\bibinfo {volume} {9}},\
  \bibinfo {pages} {340} (\bibinfo {year} {2026})}\BibitemShut {NoStop}%
\bibitem [{\citenamefont {Heinsdorf}(2025)}]{Heinsdorf2024}%
  \BibitemOpen
  \bibfield  {author} {\bibinfo {author} {\bibfnamefont {N.}~\bibnamefont
  {Heinsdorf}},\ }\href
  {https://journals.aps.org/prb/abstract/10.1103/PhysRevB.111.174407}
  {\bibfield  {journal} {\bibinfo  {journal} {Phys. Rev. B.}\ }\textbf
  {\bibinfo {volume} {111}},\ \bibinfo {pages} {174407} (\bibinfo {year}
  {2025})}\BibitemShut {NoStop}%
\bibitem [{\citenamefont {Wu}\ and\ \citenamefont {Das~Sarma}(2020)}]{Wu2020}%
  \BibitemOpen
  \bibfield  {author} {\bibinfo {author} {\bibfnamefont {F.}~\bibnamefont
  {Wu}}\ and\ \bibinfo {author} {\bibfnamefont {S.}~\bibnamefont {Das~Sarma}},\
  }\href {https://journals.aps.org/prb/abstract/10.1103/PhysRevB.102.165118}
  {\bibfield  {journal} {\bibinfo  {journal} {Phys. Rev. B Condens. Matter}\
  }\textbf {\bibinfo {volume} {102}},\ \bibinfo {pages} {165118} (\bibinfo
  {year} {2020})}\BibitemShut {NoStop}%
\bibitem [{\citenamefont {Kang}\ \emph {et~al.}(2024)\citenamefont {Kang},
  \citenamefont {Oh}, \citenamefont {Lee},\ and\ \citenamefont
  {Yang}}]{Kang2024}%
  \BibitemOpen
  \bibfield  {author} {\bibinfo {author} {\bibfnamefont {J.}~\bibnamefont
  {Kang}}, \bibinfo {author} {\bibfnamefont {T.}~\bibnamefont {Oh}}, \bibinfo
  {author} {\bibfnamefont {J.}~\bibnamefont {Lee}},\ and\ \bibinfo {author}
  {\bibfnamefont {B.-J.}\ \bibnamefont {Yang}},\ }\href
  {https://arxiv.org/abs/2402.07171} {\bibinfo {title} {Quantum geometric bound
  for saturated ferromagnetism}} (\bibinfo {year} {2024}),\ \Eprint
  {https://arxiv.org/abs/2402.07171} {arXiv:2402.07171 [cond-mat.str-el]}
  \BibitemShut {NoStop}%
\bibitem [{\citenamefont {Zhang}\ \emph {et~al.}(2026)\citenamefont {Zhang},
  \citenamefont {Wang}, \citenamefont {Balents},\ and\ \citenamefont
  {Savary}}]{Zhang2025}%
  \BibitemOpen
  \bibfield  {author} {\bibinfo {author} {\bibfnamefont {J.-X.}\ \bibnamefont
  {Zhang}}, \bibinfo {author} {\bibfnamefont {W.~O.}\ \bibnamefont {Wang}},
  \bibinfo {author} {\bibfnamefont {L.}~\bibnamefont {Balents}},\ and\ \bibinfo
  {author} {\bibfnamefont {L.}~\bibnamefont {Savary}},\ }\href
  {https://journals.aps.org/prl/abstract/10.1103/rw6g-w7my} {\bibfield
  {journal} {\bibinfo  {journal} {Phys. Rev. Lett.}\ }\textbf {\bibinfo
  {volume} {136}},\ \bibinfo {pages} {176504} (\bibinfo {year}
  {2026})}\BibitemShut {NoStop}%
\bibitem [{\citenamefont {Kitamura}\ \emph
  {et~al.}(2022{\natexlab{a}})\citenamefont {Kitamura}, \citenamefont {Daido},\
  and\ \citenamefont {Yanase}}]{Kitamura2022}%
  \BibitemOpen
  \bibfield  {author} {\bibinfo {author} {\bibfnamefont {T.}~\bibnamefont
  {Kitamura}}, \bibinfo {author} {\bibfnamefont {A.}~\bibnamefont {Daido}},\
  and\ \bibinfo {author} {\bibfnamefont {Y.}~\bibnamefont {Yanase}},\ }\href
  {https://journals.aps.org/prb/abstract/10.1103/PhysRevB.106.184507}
  {\bibfield  {journal} {\bibinfo  {journal} {Phys. Rev. B Condens. Matter}\
  }\textbf {\bibinfo {volume} {106}},\ \bibinfo {pages} {184507} (\bibinfo
  {year} {2022}{\natexlab{a}})}\BibitemShut {NoStop}%
\bibitem [{\citenamefont {Jiang}\ and\ \citenamefont
  {Barlas}(2023)}]{Jiang2023}%
  \BibitemOpen
  \bibfield  {author} {\bibinfo {author} {\bibfnamefont {G.}~\bibnamefont
  {Jiang}}\ and\ \bibinfo {author} {\bibfnamefont {Y.}~\bibnamefont {Barlas}},\
  }\href {https://journals.aps.org/prl/abstract/10.1103/PhysRevLett.131.016002}
  {\bibfield  {journal} {\bibinfo  {journal} {Phys. Rev. Lett.}\ }\textbf
  {\bibinfo {volume} {131}},\ \bibinfo {pages} {016002} (\bibinfo {year}
  {2023})}\BibitemShut {NoStop}%
\bibitem [{\citenamefont {Chen}\ and\ \citenamefont {Huang}(2023)}]{Chen2023}%
  \BibitemOpen
  \bibfield  {author} {\bibinfo {author} {\bibfnamefont {W.}~\bibnamefont
  {Chen}}\ and\ \bibinfo {author} {\bibfnamefont {W.}~\bibnamefont {Huang}},\
  }\href {https://link.springer.com/article/10.1007/s11433-023-2122-4}
  {\bibfield  {journal} {\bibinfo  {journal} {Sci. China \TK{Phys. Mech.
  Astron.}}\ }\textbf {\bibinfo {volume} {66}},\ \bibinfo {pages} {287212}
  (\bibinfo {year} {2023})}\BibitemShut {NoStop}%
\bibitem [{\citenamefont {Kitamura}\ \emph {et~al.}(2023)\citenamefont
  {Kitamura}, \citenamefont {Kanasugi}, \citenamefont {Chazono},\ and\
  \citenamefont {Yanase}}]{Kitamura2023}%
  \BibitemOpen
  \bibfield  {author} {\bibinfo {author} {\bibfnamefont {T.}~\bibnamefont
  {Kitamura}}, \bibinfo {author} {\bibfnamefont {S.}~\bibnamefont {Kanasugi}},
  \bibinfo {author} {\bibfnamefont {M.}~\bibnamefont {Chazono}},\ and\ \bibinfo
  {author} {\bibfnamefont {Y.}~\bibnamefont {Yanase}},\ }\href
  {https://journals.aps.org/prb/abstract/10.1103/PhysRevB.107.214513}
  {\bibfield  {journal} {\bibinfo  {journal} {Phys. Rev. B Condens. Matter}\
  }\textbf {\bibinfo {volume} {107}},\ \bibinfo {pages} {214513} (\bibinfo
  {year} {2023})}\BibitemShut {NoStop}%
\bibitem [{\citenamefont {Sun}\ \emph {et~al.}(2025)\citenamefont {Sun},
  \citenamefont {Yu}, \citenamefont {Chen}, \citenamefont {Hu},\ and\
  \citenamefont {Law}}]{Sun2024}%
  \BibitemOpen
  \bibfield  {author} {\bibinfo {author} {\bibfnamefont {Z.-T.}\ \bibnamefont
  {Sun}}, \bibinfo {author} {\bibfnamefont {R.-P.}\ \bibnamefont {Yu}},
  \bibinfo {author} {\bibfnamefont {S.~A.}\ \bibnamefont {Chen}}, \bibinfo
  {author} {\bibfnamefont {J.-X.}\ \bibnamefont {Hu}},\ and\ \bibinfo {author}
  {\bibfnamefont {K.~T.}\ \bibnamefont {Law}},\ }\href
  {https://link.springer.com/article/10.1007/s44214-025-00093-5} {\bibfield
  {journal} {\bibinfo  {journal} {Quantum Front.}\ }\textbf {\bibinfo {volume}
  {4}},\ \bibinfo {pages} {20} (\bibinfo {year} {2025})}\BibitemShut {NoStop}%
\bibitem [{\citenamefont {Dunbrack}\ \emph {et~al.}(2026)\citenamefont
  {Dunbrack}, \citenamefont {Virtanen},\ and\ \citenamefont
  {Heikkilä}}]{Dunbrack2025}%
  \BibitemOpen
  \bibfield  {author} {\bibinfo {author} {\bibfnamefont {A.}~\bibnamefont
  {Dunbrack}}, \bibinfo {author} {\bibfnamefont {P.}~\bibnamefont {Virtanen}},\
  and\ \bibinfo {author} {\bibfnamefont {T.~T.}\ \bibnamefont {Heikkilä}},\
  }\href {https://journals.aps.org/prb/abstract/10.1103/wnm6-rf2v} {\bibfield
  {journal} {\bibinfo  {journal} {Phys. Rev. B.}\ }\textbf {\bibinfo {volume}
  {113}},\ \bibinfo {pages} {174525} (\bibinfo {year} {2026})}\BibitemShut
  {NoStop}%
\bibitem [{\citenamefont {Shavit}\ and\ \citenamefont
  {Alicea}(2025)}]{Shavit2024}%
  \BibitemOpen
  \bibfield  {author} {\bibinfo {author} {\bibfnamefont {G.}~\bibnamefont
  {Shavit}}\ and\ \bibinfo {author} {\bibfnamefont {J.}~\bibnamefont
  {Alicea}},\ }\href {https://journals.aps.org/prb/abstract/10.1103/gt8h-czf3}
  {\bibfield  {journal} {\bibinfo  {journal} {Phys. Rev. Lett.}\ }\textbf
  {\bibinfo {volume} {134}},\ \bibinfo {pages} {176001} (\bibinfo {year}
  {2025})}\BibitemShut {NoStop}%
\bibitem [{\citenamefont {Jahin}\ and\ \citenamefont {Lin}(2026)}]{Jahin2025}%
  \BibitemOpen
  \bibfield  {author} {\bibinfo {author} {\bibfnamefont {A.}~\bibnamefont
  {Jahin}}\ and\ \bibinfo {author} {\bibfnamefont {S.-Z.}\ \bibnamefont
  {Lin}},\ }\href {https://journals.aps.org/prb/abstract/10.1103/gt8h-czf3}
  {\bibfield  {journal} {\bibinfo  {journal} {Phys. Rev. B.}\ }\textbf
  {\bibinfo {volume} {113}},\ \bibinfo {pages} {014504} (\bibinfo {year}
  {2026})}\BibitemShut {NoStop}%
\bibitem [{sym()}]{symmetry_comment}%
  \BibitemOpen
  \href@noop {} {}\bibinfo {note} {More precisely, we assume that the two-fold
  band degeneracy is enforced because two states transform according to a 2D
  irreducible representation due to the combination of $C_4$ symmetry and other
  symmetries, such as mirror or spinless time-reversal symmetries.}\BibitemShut
  {Stop}%
\bibitem [{App()}]{Appendix}%
  \BibitemOpen
  \href@noop {} {}\bibinfo {note} {See End Matter for more
  details.}\BibitemShut {Stop}%
\bibitem [{T_d()}]{T_dive_comment}%
  \BibitemOpen
  \href@noop {} {}\bibinfo {note} {The $1/T$ divergence of $\chi_{\rm c}$ has
  been pointed out in Ref.~\cite{Kitamura2024}.}\BibitemShut {Stop}%
\bibitem [{\citenamefont {Nogaki}\ and\ \citenamefont
  {Yanase}(2024)}]{Nogaki2024}%
  \BibitemOpen
  \bibfield  {author} {\bibinfo {author} {\bibfnamefont {K.}~\bibnamefont
  {Nogaki}}\ and\ \bibinfo {author} {\bibfnamefont {Y.}~\bibnamefont
  {Yanase}},\ }\href {https://doi.org/10.1103/PhysRevB.110.184501} {\bibfield
  {journal} {\bibinfo  {journal} {Phys. Rev. B}\ }\textbf {\bibinfo {volume}
  {110}},\ \bibinfo {pages} {184501} (\bibinfo {year} {2024})}\BibitemShut
  {NoStop}%
\bibitem [{\citenamefont {Raghu}\ \emph {et~al.}(2008)\citenamefont {Raghu},
  \citenamefont {Qi}, \citenamefont {Liu}, \citenamefont {Scalapino},\ and\
  \citenamefont {Zhang}}]{Raghu2008}%
  \BibitemOpen
  \bibfield  {author} {\bibinfo {author} {\bibfnamefont {S.}~\bibnamefont
  {Raghu}}, \bibinfo {author} {\bibfnamefont {X.-L.}\ \bibnamefont {Qi}},
  \bibinfo {author} {\bibfnamefont {C.-X.}\ \bibnamefont {Liu}}, \bibinfo
  {author} {\bibfnamefont {D.~J.}\ \bibnamefont {Scalapino}},\ and\ \bibinfo
  {author} {\bibfnamefont {S.-C.}\ \bibnamefont {Zhang}},\ }\href
  {https://journals.aps.org/prb/abstract/10.1103/PhysRevB.77.220503} {\bibfield
   {journal} {\bibinfo  {journal} {Phys. Rev. B Condens. Matter}\ }\textbf
  {\bibinfo {volume} {77}},\ \bibinfo {pages} {220503} (\bibinfo {year}
  {2008})}\BibitemShut {NoStop}%
\bibitem [{\citenamefont {Bergman}\ \emph {et~al.}(2008)\citenamefont
  {Bergman}, \citenamefont {Wu},\ and\ \citenamefont
  {Balents}}]{Bergman2008-nc}%
  \BibitemOpen
  \bibfield  {author} {\bibinfo {author} {\bibfnamefont {D.~L.}\ \bibnamefont
  {Bergman}}, \bibinfo {author} {\bibfnamefont {C.}~\bibnamefont {Wu}},\ and\
  \bibinfo {author} {\bibfnamefont {L.}~\bibnamefont {Balents}},\ }\href
  {https://journals.aps.org/prb/abstract/10.1103/PhysRevB.78.125104} {\bibfield
   {journal} {\bibinfo  {journal} {Phys. Rev. B Condens. Matter Mater. Phys.}\
  }\textbf {\bibinfo {volume} {78}},\ \bibinfo {pages} {125104} (\bibinfo
  {year} {2008})}\BibitemShut {NoStop}%
\bibitem [{\citenamefont {Rhim}\ and\ \citenamefont {Yang}(2019)}]{Rhim2019}%
  \BibitemOpen
  \bibfield  {author} {\bibinfo {author} {\bibfnamefont {J.-W.}\ \bibnamefont
  {Rhim}}\ and\ \bibinfo {author} {\bibfnamefont {B.-J.}\ \bibnamefont
  {Yang}},\ }\href
  {https://journals.aps.org/prb/abstract/10.1103/PhysRevB.99.045107} {\bibfield
   {journal} {\bibinfo  {journal} {Phys. Rev. B Condens. Matter}\ }\textbf
  {\bibinfo {volume} {99}},\ \bibinfo {pages} {045107} (\bibinfo {year}
  {2019})}\BibitemShut {NoStop}%
\bibitem [{\citenamefont {Marzari}\ and\ \citenamefont
  {Vanderbilt}(1997)}]{Marzari1997}%
  \BibitemOpen
  \bibfield  {author} {\bibinfo {author} {\bibfnamefont {N.}~\bibnamefont
  {Marzari}}\ and\ \bibinfo {author} {\bibfnamefont {D.}~\bibnamefont
  {Vanderbilt}},\ }\href
  {https://journals.aps.org/prb/abstract/10.1103/PhysRevB.56.12847} {\bibfield
  {journal} {\bibinfo  {journal} {Phys. Rev. B Condens. Matter}\ }\textbf
  {\bibinfo {volume} {56}},\ \bibinfo {pages} {12847} (\bibinfo {year}
  {1997})}\BibitemShut {NoStop}%
\bibitem [{\citenamefont {Vanderbilt}()}]{Vanderbilt2018}%
  \BibitemOpen
  \bibfield  {author} {\bibinfo {author} {\bibfnamefont {D.}~\bibnamefont
  {Vanderbilt}},\ }\href {https://doi.org/10.1017/9781316662205} {\bibinfo
  {title} {Berry phases in electronic structure theory: Electric polarization,
  orbital magnetization and topological insulators}},\ \bibinfo {note}
  {(Cambridge University Press, 2018)}\BibitemShut {NoStop}%
\bibitem [{\citenamefont {Hu}\ \emph {et~al.}(2025)\citenamefont {Hu},
  \citenamefont {Liang}, \citenamefont {Li}, \citenamefont {Li}, \citenamefont
  {Meng}, \citenamefont {Lei}, \citenamefont {Wang}, \citenamefont {Wen},
  \citenamefont {Zhang}, \citenamefont {Cai}, \citenamefont {Zhang},
  \citenamefont {Lu}, \citenamefont {Wang}, \citenamefont {Xue},\ and\
  \citenamefont {Zhang}}]{Hu2025-dt}%
  \BibitemOpen
  \bibfield  {author} {\bibinfo {author} {\bibfnamefont {Y.}~\bibnamefont
  {Hu}}, \bibinfo {author} {\bibfnamefont {K.}~\bibnamefont {Liang}}, \bibinfo
  {author} {\bibfnamefont {J.}~\bibnamefont {Li}}, \bibinfo {author}
  {\bibfnamefont {Z.}~\bibnamefont {Li}}, \bibinfo {author} {\bibfnamefont
  {F.}~\bibnamefont {Meng}}, \bibinfo {author} {\bibfnamefont {H.}~\bibnamefont
  {Lei}}, \bibinfo {author} {\bibfnamefont {J.}~\bibnamefont {Wang}}, \bibinfo
  {author} {\bibfnamefont {H.}~\bibnamefont {Wen}}, \bibinfo {author}
  {\bibfnamefont {R.}~\bibnamefont {Zhang}}, \bibinfo {author} {\bibfnamefont
  {J.}~\bibnamefont {Cai}}, \bibinfo {author} {\bibfnamefont {J.}~\bibnamefont
  {Zhang}}, \bibinfo {author} {\bibfnamefont {Y.}~\bibnamefont {Lu}}, \bibinfo
  {author} {\bibfnamefont {Y.}~\bibnamefont {Wang}}, \bibinfo {author}
  {\bibfnamefont {Q.-K.}\ \bibnamefont {Xue}},\ and\ \bibinfo {author}
  {\bibfnamefont {D.}~\bibnamefont {Zhang}},\ }\href
  {https://www.nature.com/articles/s41467-025-62624-x} {\bibfield  {journal}
  {\bibinfo  {journal} {Nat. Commun.}\ }\textbf {\bibinfo {volume} {16}},\
  \bibinfo {pages} {7305} (\bibinfo {year} {2025})}\BibitemShut {NoStop}%
\bibitem [{\citenamefont {Kitamura}\ \emph {et~al.}(2021)\citenamefont
  {Kitamura}, \citenamefont {Ishizuka}, \citenamefont {Daido},\ and\
  \citenamefont {Yanase}}]{Kitamura2021-jk}%
  \BibitemOpen
  \bibfield  {author} {\bibinfo {author} {\bibfnamefont {T.}~\bibnamefont
  {Kitamura}}, \bibinfo {author} {\bibfnamefont {J.}~\bibnamefont {Ishizuka}},
  \bibinfo {author} {\bibfnamefont {A.}~\bibnamefont {Daido}},\ and\ \bibinfo
  {author} {\bibfnamefont {Y.}~\bibnamefont {Yanase}},\ }\href
  {https://journals.aps.org/prb/abstract/10.1103/PhysRevB.103.245114}
  {\bibfield  {journal} {\bibinfo  {journal} {Phys. Rev. B Condens. Matter}\
  }\textbf {\bibinfo {volume} {103}},\ \bibinfo {pages} {245114} (\bibinfo
  {year} {2021})}\BibitemShut {NoStop}%
\bibitem [{\citenamefont {Kitamura}\ \emph
  {et~al.}(2022{\natexlab{b}})\citenamefont {Kitamura}, \citenamefont
  {Yamashita}, \citenamefont {Ishizuka}, \citenamefont {Daido},\ and\
  \citenamefont {Yanase}}]{Kitamura2022-wi}%
  \BibitemOpen
  \bibfield  {author} {\bibinfo {author} {\bibfnamefont {T.}~\bibnamefont
  {Kitamura}}, \bibinfo {author} {\bibfnamefont {T.}~\bibnamefont {Yamashita}},
  \bibinfo {author} {\bibfnamefont {J.}~\bibnamefont {Ishizuka}}, \bibinfo
  {author} {\bibfnamefont {A.}~\bibnamefont {Daido}},\ and\ \bibinfo {author}
  {\bibfnamefont {Y.}~\bibnamefont {Yanase}},\ }\href
  {https://journals.aps.org/prresearch/abstract/10.1103/PhysRevResearch.4.023232}
  {\bibfield  {journal} {\bibinfo  {journal} {Phys. Rev. Research}\ }\textbf
  {\bibinfo {volume} {4}},\ \bibinfo {pages} {023232} (\bibinfo {year}
  {2022}{\natexlab{b}})}\BibitemShut {NoStop}%
\bibitem [{Sup()}]{Supple}%
  \BibitemOpen
  \href@noop {} {}\bibinfo {note} {See the Supplemental Material for details on
  the singular saddle point in the kagome lattice model and the construction of
  the quasi-local states in the $t_{2g}$-orbital model.}\BibitemShut {Stop}%
\end{thebibliography}

\begin{thebibliography}{4}%
\makeatletter
\providecommand \@ifxundefined [1]{%
 \@ifx{#1\undefined}
}%
\providecommand \@ifnum [1]{%
 \ifnum #1\expandafter \@firstoftwo
 \else \expandafter \@secondoftwo
 \fi
}%
\providecommand \@ifx [1]{%
 \ifx #1\expandafter \@firstoftwo
 \else \expandafter \@secondoftwo
 \fi
}%
\providecommand \natexlab [1]{#1}%
\providecommand \enquote  [1]{``#1''}%
\providecommand \bibnamefont  [1]{#1}%
\providecommand \bibfnamefont [1]{#1}%
\providecommand \citenamefont [1]{#1}%
\providecommand \href@noop [0]{\@secondoftwo}%
\providecommand \href [0]{\begingroup \@sanitize@url \@href}%
\providecommand \@href[1]{\@@startlink{#1}\@@href}%
\providecommand \@@href[1]{\endgroup#1\@@endlink}%
\providecommand \@sanitize@url [0]{\catcode `\\12\catcode `\$12\catcode
  `\&12\catcode `\#12\catcode `\^12\catcode `\_12\catcode `\%12\relax}%
\providecommand \@@startlink[1]{}%
\providecommand \@@endlink[0]{}%
\providecommand \url  [0]{\begingroup\@sanitize@url \@url }%
\providecommand \@url [1]{\endgroup\@href {#1}{\urlprefix }}%
\providecommand \urlprefix  [0]{URL }%
\providecommand \Eprint [0]{\href }%
\providecommand \doibase [0]{https://doi.org/}%
\providecommand \selectlanguage [0]{\@gobble}%
\providecommand \bibinfo  [0]{\@secondoftwo}%
\providecommand \bibfield  [0]{\@secondoftwo}%
\providecommand \translation [1]{[#1]}%
\providecommand \BibitemOpen [0]{}%
\providecommand \bibitemStop [0]{}%
\providecommand \bibitemNoStop [0]{.\EOS\space}%
\providecommand \EOS [0]{\spacefactor3000\relax}%
\providecommand \BibitemShut  [1]{\csname bibitem#1\endcsname}%
\let\auto@bib@innerbib\@empty
\bibitem [{\citenamefont {Bergman}\ \emph {et~al.}(2008)\citenamefont
  {Bergman}, \citenamefont {Wu},\ and\ \citenamefont
  {Balents}}]{SBergman2008-nc}%
  \BibitemOpen
  \bibfield  {author} {\bibinfo {author} {\bibfnamefont {D.~L.}\ \bibnamefont
  {Bergman}}, \bibinfo {author} {\bibfnamefont {C.}~\bibnamefont {Wu}},\ and\
  \bibinfo {author} {\bibfnamefont {L.}~\bibnamefont {Balents}},\ }\href
  {https://journals.aps.org/prb/abstract/10.1103/PhysRevB.78.125104} {\bibfield
   {journal} {\bibinfo  {journal} {Phys. Rev. B Condens. Matter Mater. Phys.}\
  }\textbf {\bibinfo {volume} {78}},\ \bibinfo {pages} {125104} (\bibinfo
  {year} {2008})}\BibitemShut {NoStop}%
\bibitem [{\citenamefont {Rhim}\ and\ \citenamefont
  {Yang}(2019)}]{SRhim2019-km}%
  \BibitemOpen
  \bibfield  {author} {\bibinfo {author} {\bibfnamefont {J.-W.}\ \bibnamefont
  {Rhim}}\ and\ \bibinfo {author} {\bibfnamefont {B.-J.}\ \bibnamefont
  {Yang}},\ }\href
  {https://journals.aps.org/prb/abstract/10.1103/PhysRevB.99.045107} {\bibfield
   {journal} {\bibinfo  {journal} {Phys. Rev. B Condens. Matter}\ }\textbf
  {\bibinfo {volume} {99}},\ \bibinfo {pages} {045107} (\bibinfo {year}
  {2019})}\BibitemShut {NoStop}%
\bibitem [{\citenamefont {Mielke}(1999)}]{SMielke1999}%
  \BibitemOpen
  \bibfield  {author} {\bibinfo {author} {\bibfnamefont {A.}~\bibnamefont
  {Mielke}},\ }\href
  {https://iopscience.iop.org/article/10.1088/0305-4470/32/48/304} {\bibfield
  {journal} {\bibinfo  {journal} {J. Phys. A Math. Gen.}\ }\textbf {\bibinfo
  {volume} {32}},\ \bibinfo {pages} {8411} (\bibinfo {year}
  {1999})}\BibitemShut {NoStop}%
\bibitem [{\citenamefont {Tasaki}()}]{STasaki2020}%
  \BibitemOpen
  \bibfield  {author} {\bibinfo {author} {\bibfnamefont {H.}~\bibnamefont
  {Tasaki}},\ }\href
  {https://doi.org/https://doi.org/10.1007/978-3-030-41265-4} {\bibinfo {title}
  {Physics and mathematics of quantum many-body systems}},\ \bibinfo {note}
  {(Springer, Cham, 2020)}\BibitemShut {NoStop}%
\end{thebibliography}
\end{document}